\documentclass[preprint,12pt]{elsarticle}
\usepackage[margin=2.5cm]{geometry}
\usepackage[utf8]{inputenc}
\usepackage{xcolor}
\usepackage{bm}
\usepackage{graphicx}
\usepackage{verbatim}
\usepackage{soul}
\usepackage{multirow}
\usepackage{natbib}
\setcitestyle{square, comma, numbers,sort&compress}
\usepackage{bbm}

\usepackage[commandnameprefix=always,todonotes={textsize=tiny}]{changes}
\usepackage{algorithm}
\usepackage{algorithmic}
\usepackage{appendix}
\usepackage{subcaption}
\usepackage{setspace}
\usepackage{booktabs}
\usepackage{amsmath}
\numberwithin{equation}{section}
\usepackage[makeroom]{cancel}
\newcommand{\stkout}[1]{\ifmmode\text{\sout{\ensuremath{#1}}}\else\sout{#1}\fi}
\usepackage{amssymb}
\usepackage[colorlinks]{hyperref} 
\hypersetup{ 
    colorlinks=true,       % false: boxed links; true: colored links
    linkcolor=red,          % color of internal links
    citecolor=blue,        % color of links to bibliography
    filecolor=magenta,      % color of file links
    urlcolor=cyan           % color of external links
}

\usepackage{tikz}
\usepackage{stmaryrd}

\newcommand{\bx}{\bm{x}}

\newcommand{\fref}[1]{Fig.~\ref{#1}}

\newcommand{\eref}[1]{(\ref{#1})}

\newcommand{\sref}[1]{Section~\ref{#1}}

\newcommand{\blue}[1]{\color{blue}}{}

\journal{Modelling and Simulation in Materials Science and Engineering}

\begin{document}

\begin{frontmatter}

\title{Statistics of grain microstructure evolution under anisotropic grain boundary energies and mobilities using threshold-dynamics}

%\author{Names of authors are omitted for a double-anonymous peer review process}
% MSMS holds a double-anoymous review, so I am deleting the author information now.  

\address[mechseaddress]{Department of Mechanical Science and Engineering, University of Illinois Urbana--Champaign, United States}
\address[hongikaddress]{Department of Mechanical and Design Engineering, Hongik University, Sejong, South Korea}

\author[mechseaddress,hongikaddress]{Jaekwang Kim}
\author[mechseaddress]{Nikhil Chandra Admal}

\begin{abstract}
This paper investigates the statistical behavior of two-dimensional grain microstructures during grain growth under anisotropic grain boundary characters. We employ the threshold-dynamics method, which allows for unparalleled computational speed, to simulate the full-field curvature motion of grain boundaries in a large polycrystal ensemble. Two sets of numerical experiments are performed to explore the effect of grain boundary anisotropy on the evolution of microstructure features.
In the first experiment, we focus on abnormal grain growth and find that grain boundary anisotropy introduces a statistical preference for certain grain orientations. This leads to changes in the overall grain size distribution from the isotropic case. In the second experiment, we examine the texture development and growth of twin grain boundaries at different initial microstructures. We find that both phenomena are more pronounced when the initial microstructure has a dominant fraction of high-angle grain boundaries.
Our results suggest effective grain boundary engineering strategies for improving material properties.
\end{abstract}
\end{frontmatter}

\section{Introduction}

The macroscopic properties of polycrystalline materials are strongly
influenced by their grain microstructure, which is determined by the thermomechanical
loads during materials processing. At the macroscale, a grain microstructure is typically
described by the grain boundary (GB) character distribution. GB engineering
refers to the strategy of enhancing the properties of a polycrystal
by transforming its GB character distribution to a target
distribution using thermomechanical processes~\cite{Watanabe:2011}.
Recently, the GB engineering paradigm has been extended to tailor
the properties of nanocrystalline materials, which are promising next-generation
structural materials with high strength, fatigue life, and wear
resistance~\citep{NCmaterial01,NCmaterial02,NCmaterial03}. For example, subjecting nanocrystalline materials to thermomechanical cycling leads to a  
considerable increase in the fraction of $\Sigma 3$ grain
boundaries~\citep{Rupert:2015}, which demonstrate high resistance to sliding,
cavitation, and fracture~\citep{NCmaterial04}. 

Establishing the relationship between microstructure and materials properties, as
well as predicting the evolution of microstructure during various manufacturing
processes are ongoing research challenges in the field of GB
engineering. In this regard,
atomistic~\citep{Frolov:2018,Foiles:2006,MOLODOV2018336},
mesoscopic~\citep{Srolovitz:2017, JKIM:2021}, and macroscopic
continuum~\citep{Belytschko:2009,ZHAO2022275,LIU2014310} models have been
developed over the last few decades to uncover the connection between microstructures, properties, and process parameters.  

A defining characteristic of these models is the motion of GBs
driven by surface tension to decrease the interfacial energy. 
When a polycrystalline material is annealed,
grains grow to decrease the total energy of the system by reducing the
GB area, leading to the motion of GBs towards their centers of
curvatures. The Mullins model~\citep{Mullins} describes such a motion as
\begin{equation}
v=-m \gamma \kappa, 
\label{e:motionbycurvature}
\end{equation}
where $v$, $\kappa$, $\gamma$, and $m$  denote signed velocity magnitude, curvature, misorientation-dependent energy
density, and mobility of the GB, respectively.
If $m$ and $\gamma$ are constants, the system is called
\emph{isotropic}. Grain growth in an isotropic system is characterized by a relatively
narrow range in the grain size distributions that obeys a simple scaling
relation, regardless of the initial configuration of the grain microstructure~\citep{Barmak:2013,Lazar:2020}.
On the other hand, anisotropic grain growth occurs
when $m$ and $\gamma$ depend on the GB character, defined by the
five macroscopic degrees of freedom (dofs).\footnote{Here, three degrees
represent a rotation associated with the misorientation between the two grains,
and the remaining two degrees correspond to the inclination of the grain
boundary.} 

Recent studies have reported that microstructure evolution 
is heavily influenced by the anisotropy of GBs~\citep{JKIM:2021,Mohles,BLEE:2014,SALAMA2020641}.
With the aid of accurate interatomic potentials, significant progress has been
made in mapping the GB anisotropy by the construction of grain
boundary energies and mobilities as functions of the GB
character~\citep{Srolovitz:2020_2, Runnels:2016_1,
Runnels:2016_2, Olmsted:2009,Bulatov:2019}. However, precisely identifying the GB character distribution 
responsible for spontaneous microstructure transformation phenomena such as
\textit{abnormal grain growth} (AGG) and \textit{recrystallization}
remains a fundamental challenge in materials science. 
One of the main difficulties arises from the enormity of the microstructure
space (relative to the space of processes and properties), which
encompasses all possible configurations of grain assemblies. Due to the
relatively small size of the property space, the structure-property relationship is necessarily a many-to-one
mapping~\citep{kalidindi2015materials,kalidindi2015hierarchical}. Therefore,
grain microstructures are commonly described using statistical distributions for
grain sizes, topologies, and orientations, which makes it possible to express
structure-property relationship using reduced order models~\citep{JKIM:2023,ABBRUZZESE1986905,Pande:2001}.
For example, the reduced order model developed by \citet{JKIM:2023} describes the evolution of distributions for grain sizes and topologies under the idealized isotropic grain growth. 
Extending the model to anisotropic grain growth requires careful
investigation of the role of GB anisotropy on the evolution of the
statistics of a microstructure, which is the main goal of this paper.

\textit{Threshold-dynamics} (TD)~\citep{Esedoglu:2015}, a class of numerical methods to simulate GB motion, has proven to be highly efficient in simulating large ensembles of
grains.
In the TD framework, GBs are sharp interfaces 
that evolve according to motion by curvature, given in~\eref{e:motionbycurvature}.
Each grain is described by a level set function that is assigned a positive
value within the grain and a negative value outside of it. This implies that the
zero-valued isosurface of the function corresponds to the interface surrounding
the grain. One advantage of the sharp interface framework is that it
allows a relatively coarse grid representation of a GB compared to diffuse-interface models (including
phase-field models), which require the grid to be refined enough to resolve the
width of a GB. In addition, the high computational efficiency of the TD
methods stems from the fast Fourier transform (FFT) algorithms used to implement the diffusion operator,
which drives GB motion. 
In addition, the TD scheme was recently equipped with a variational structure for multi-phase settings~\citep{Esedoglu:2015}. 
This ensured the correct prediction of the dihedral angles between three anisotropic
interfaces at triple junctions, which would necessitate considerably more
computational effort to achieve using diffused-interface
models~\citep{JKIM:2021,Mason:2022,Dziwnik:2017}.
The superior computational performance of the TD methods has enabled
simulations of large collections of grains that adequately sample microstructure
distributions. \citet{Esedoglu:2019} used the TD method to investigate the
evolution of grain size and topology distributions in two dimensions and
compared them to the predictions of experiments and the phase-field-crystal method.~\footnote{
The phase field crystal method resolves the atomic positions and describes their evolution at a diffusive time scale. This technique involves a free energy functional that is minimized when the density field is periodic, thereby facilitating the formation of density field patterns in solid phases~\citep{PFC}. 
} 
\citet{Rohrer:2022} examined the
evolution of grain morphologies of individual grains of Ni in three dimensions and the
overall grain size distribution, and compared to experiments.  However, their
implementation of TD was limited to an isotropic system with constant GB
energy and mobility. GB anisotropy was noted as a contributing factor in
instances when disagreement with experiments was identified.
On the other hand, \citet{NINO2023111879} implemented the TD scheme for anisotropic systems using several GB energy functions, including the energy function 
of \citet{BULATOV2014161} that depends on all five dofs of a GB.
While they observed distinct signatures in the morphological evolution of
individual grains for different GB energy functions, GB statistics were reported to evolve identically in all the cases. 
However, it must be noted that \citet{NINO2023111879} assumed 
a unit reduced mobility, 
meaning that the product of the mobility and surface tension is set to one. 
In this case, although the dihedral angles at triple junctions are dictated by GB energy anisotropy, the anisotropy of surface tension is canceled by that of 
mobility. Consequently, the resulting motion by curvature remains similar to the isotropic case.
In our view, it is important to isolate GB mobility to appropriately estimate the role of GB
energy anisotropy. Fortunately, this is made possible by a recent update to the
algorithm of \citet{Salvador:2019}, which allows
the prescription of GB energies and mobilities independently.
Using the latest TD algorithm, \citet{Salvador:2019_2} investigated grain
statistics under a constant mobility and a
Read--Shockley GB energy (RSE). \footnote{A grain boundary energy is of the Read--Shockley form, if it is a monotonously increasing concave function of the misorientation angle.  
} 
It was observed that in two dimensions, 
the RSE also did not lead to significant deviation in the grain size distribution compared to the isotropic case.  
However, since the RSE is valid for only small misorientation angles, the full extent
of the impact of GB energy anisotropy on grain growth remains unexplored.

In this paper, we investigate the influence of GB anisotropy on the statistical
evolution of grain microstructure and its steady state in
two dimensions. Our first objective is to investigate how GB energy anisotropy affects
AGG, a key feature of microstructure evolution that
significantly impacts materials' properties. 
Secondly, we will investigate texture development and the growth of special boundaries 
in initial microstructure configurations with different fractions of 
low-angle grain boundaries (LAGBs) and high-angle grain boundaries (HAGBs).  
The question is particularly significant in GB engineering of nanocrystalline materials. 
For example, \textit{Equal Channel Angular Pressing} (ECAP)~\citep{ECAP:2000,ECAP:2004},
which is a popular synthesizing nanostructured solids,   
involves severe plastic deformation of coarse-grained materials
leading to substantial grain refinement and a nanostructure~\citep{Langdon}.
While it has been reported that the number of ECAP cycles largely determines
the fraction of HAGB and LAGB~\citep{ECAP:2020}, it is still not clear how these
different initial microstructures evolve during subsequent annealing
treatment. Our study aims to shed light on this important question.

The manuscript is organized as follows. \sref{sec:method} provides a brief
introduction to the TD with anisotropic GB energy and mobility, and its implementation that
ensures numerical stability. In \sref{sec:simulations}, we present results of carefully
designed numerical experiments to achieve the goals of this paper.
We summarize and conclude in \sref{sec:conclusion}. Table~\ref{tab:abbreviations}
collects the abbreviations used in this paper.
\begin{table}[h]
\begin{center}
\begin{tabular}{lll}
\textbf{Abbreviation} & \textbf{Defintion} \\
GB & grain boundary   \\
LAGB & low-angle grain boundary \\
HAGB & high-angle grain boundary \\
TB & twin boundary \\
HAGB* & high angle grain boundaries excluding twin boundaries \\
STGB & symmetric tilt grain boundary \\
RSE & Read--Shockley (grain boundary) energy \\
AGG & abnormal grain growth \\
TD & threshold-dynamics \\
PSP & process-(micro)structure-property \\
ECAP & equal channel angular pressing
\end{tabular}
\caption{List of abbreviations}
\label{tab:abbreviations}
\end{center}
\end{table}

\section{Method}
\label{sec:method}

The TD algorithm employs two highly efficient operations in an alternating
manner to simulate motion by curvature. The first operation is a convolution of
a radially symmetric kernel with a characteristic function, which is equal to 1 inside the
interface and 0 outside. In the second operation, the resulting output is
subjected to point-wise \textit{thresholding} to obtain an updated characteristic function.
The TD method was originally introduced by Merriman, Bence, and Osher
(MBO)~\citep{MBO} for a two-phase system. The key idea is that the level-set of a distance function, under the action of a diffusion operator, moves in the normal direction with a
velocity equal to the mean curvature of the level-set surface. 
While many extensions to multi-phase systems were
subsequently developed, the generalization of~\citet{Esedoglu:2015} demonstrates
superior characteristics due to its variational structure.
Moreover, recent developments to the TD method include enhanced
accuracy~\citep{ZAITZEFF2020109404}, grain grouping techniques to save
computational memory~\citep{Esedoglu:2009,Esedoglu:2011}, and incorporating anisotropic
mobility~\citep{ESEDOLU201762}.
In the following section, we summarize the TD method used in this paper.

\subsection{Background}
Consider a two-dimensional polycrystal $D=\cup^N_{j=1}\Sigma_j$ partitioned by 
$N$ grains occupying regions $\Sigma_j$. If $\gamma_{ij}(=\gamma_{ji})$ denotes the
interfacial energy density between grains $\Sigma_i$ and $\Sigma_j$, the total
grain boundary energy of the polycrystal is given by 
\begin{equation}
    E=\frac{1}{2}\sum_{i,j=1}^N \gamma_{ij}\mathrm{Area}(\Gamma_{ij}),
    \label{e:general_energy}
\end{equation}
where $\Gamma_{ij}$ denotes the boundary between two adjacent grains $\Sigma_i$
and $\Sigma_j$.
If $\gamma$ represents the matrix with $\gamma_{ij}$ as its entries,
then $\gamma$ belongs to the following class of surface tension matrices
\begin{equation}
    \mathcal S _N =\{
\gamma \in R^{N\times N}: \gamma_{ii}=0 \quad
\mathrm{and} \quad
\gamma_{ij}=\gamma_{ji}>0 \quad
\mathrm{for \; all \;distinct\;} i, j 
\}.
\label{e:general_surfacetension}
\end{equation}

The steepest descent of the energy in \eqref{e:general_energy} results in an
anisotropic motion by curvature,
\begin{equation*}
v_{ij}=-\mu_{ij}\gamma_{ij} \kappa_{ij},
\end{equation*}
where $v_{ij}$, $\kappa_{ij}$, and $\mu_{ij}$ denote the velocity, mean curvature,
mobility, and mobility of the interface $\Gamma_{ij}$. 
The variational structure of the TD originates from the following non-local approximation of $E$
\begin{equation}
    E_{\delta t}(\Sigma_1, ...,\Sigma_N)= \frac{1}{2 \delta t} \sum^N_{i,j=1}\gamma_{ij} \int_D \mathbb{I}_{\Sigma_i} K_{\delta t} * \mathbb{I}_{\Sigma_j} \; dx,
    \label{e:threshold_energy}
\end{equation}
where $K_{\delta t}:\mathbbm{R}^d \to \mathbbm{R}$ is a positive-valued
convolution kernel with a characteristic width $\delta t$, and
$\mathbb{I}_{\Sigma_i}$ is the characteristic function
\begin{equation*}
   \mathbb{I}_{\Sigma_i}(\bx) =
     \begin{cases}
        1 & \text{ if }
        \bx \in \Sigma_i, \\
        0 & \text{ otherwise.}
    \end{cases}
\end{equation*}
\citet{Esedoglu:2015} showed that $E_{\delta t} \to E$, in the sense of
$\Gamma$-convergence, as $\delta t \to 0$.
$E_{\delta t}$ may be viewed as a functional of $N$ characteristic functions.
More formally, $E_{\delta t}$ is defined on the following set $\mathcal B$
consisting of $n$-tuples of binary functions 
\begin{align}
    \mathcal{B} = \{ (u_1,\dots,u_N) : &\text{ for each } \bm x \in D
    \text{ there is an } i \text{ such that } u_i(\bx)=1 \notag \\
    &\text{ and } u_j(\bx)=0
    \text{ for all }j\neq i \}. 
\label{e:characteristic_binary_space}
\end{align}
The construction of $E_{\delta t}$~\eref{e:threshold_energy} relies on the
interpretation~\citep{Esedoglu:2015} that the
surface area of $\Gamma_{ij}$ can be estimated by the amount of heat that
escapes from $\Sigma_i$ into $\Sigma_j$ in $\delta t$ time, i.e.,
\begin{equation}
    \mathrm{Area}(\Gamma_{ij}) \approx \frac{1}{\delta t} \sum^N_{i,j=1}\int_D
    \mathbb{I}_{\Sigma_i} K_{\delta t} * \mathbb{I}_{\Sigma_j} \; dx. 
\end{equation}
While a common choice of $K_{\delta t}$ is the Gaussian kernel
\begin{equation}
    G_{\delta t }(\bx)= \frac{1}{ (4\pi \delta t)^{\frac{d}{2}}} \exp \left( -\frac{ |\bx|^2}{4\delta t}\right),
    \label{e:GaussianKernel}
\end{equation}
alternate kernels have been proposed more recently.

A point-wise thresholding rule is designed to move grain boundaries in their
normal direction by distances of $\gamma_{ij}\mu_{ij}\kappa_{ij}\delta t$ at
every time step, such that the energy $E_{\delta t}$ monotonically
decreases. The characteristic width, $\delta t$, of the kernel corresponds to the time step size $t_S$. 
The parameter $\delta t$ also determines the minimum grid size $\delta x <
\gamma_{ij}\mu_{ij}\kappa_{\mathrm{min}}\delta t$ 
required by the TD algorithm, where $\kappa_{\mathrm{min}}$ is the minimum curvature (such as the curvature of the largest grain in $D$).
If the grid size condition is not satisfied,
grain boundaries would stagnate.
The variational nature of the TD algorithms
ensures the correct equilibrium dihedral angle condition at triple junctions (known as Herring angle condition~\citep{Herring}) is satisfied~\citep{Esedoglu:2015}.

\subsection{Algorithm}

The TD scheme employed in this paper is a recent version proposed by \citet{Salvador:2019}.
Comparing to the original version~\citep{Esedoglu:2015}, 
the current TD scheme broadens the choice of anisotropic mobilities $\mu_{ij}$ 
with minimum algorithmic complication. 
This is achieved by constructing the kernel $K_{\delta t}$ using two Gaussian
kernels with distinct non-negative width parameters $\alpha$ and $\beta$ as
\begin{equation}
    K_{\delta t}= a_{ij} G_{\sqrt{\alpha \delta t}} + b_{ij} G_{\sqrt{\beta \delta t}},
    \label{e:algorithm_kernel}
\end{equation}
where 
\begin{equation}
    a_{ij}=\frac{\sqrt{\pi \alpha}}{\alpha - \beta} (\gamma_{ij}-\beta \mu^{-1}_{ij}),
    \label{e:kernel_coeff_a}
\end{equation}
and 
\begin{equation}
    b_{ij}=\frac{\sqrt{\pi \beta}}{\alpha -\beta}(-\gamma_{ij}+ \alpha \mu^{-1}_{ij}).
    \label{e:kernel_coeff_b}
\end{equation}
The evolution of grain domains $\Sigma_1^n,\dots,\Sigma_N^n$ at the $n$-th step involves three steps. First, convolutions $\phi^n_{1,i}
:= G_{\sqrt{\alpha \delta t}}* \mathbb{I}_{\Sigma_i^n}$ and $\phi^n_{2,i} :=
G_{\sqrt{\beta \delta t}}* \mathbb{I}_{\Sigma_i^n}$
are computed. Second, comparison functions $\psi^n_{i}:= \sum_{i\neq
j}a_{ij}\phi^n_{1,j} + b_{ij}\phi^n_{2,j}$ are assembled. In the final step, thresholding is
carried out using the criterion
\begin{equation}
\Sigma^{n+1}_i =\{ \bx:\psi^n_i (\bx) < 
\displaystyle{
\min_{j\neq i } }\; \psi^n_j (\bx) \}.
\label{e:threshold_rule}
\end{equation}
The aforementioned steps are summarized in Algorithm~\ref{algo:salvador}.

In real materials, the dependence of GB energies and mobilities on the misorientation is highly complex with multiple local maxima/minima.
However, Algorithm~\ref{algo:salvador} and other TD versions are 
restricted to the following class of GB energies 
\begin{equation}
 \mathcal{T}_N =\{\gamma \in \mathcal{S}_N:
 \gamma_{ij} + \gamma_{jk} \leq \gamma_{ik} \quad
 \mathrm{for \; any \;}i,j,k 
 \},
 \label{e:triangular_inequality}
\end{equation}
which includes energies that satisfy a triangle inequality. If $\gamma \notin
\mathcal{T_N}$, a TD scheme may lead to grain boundary wetting by nucleating a
new grain along one of the existing boundaries~\citep{Esedoglu:2015}.
Such a nucleation is entirely a numerical artifact and thus unphysical.
Wetting can be circumvented by restricting the thresholding condition in
\eref{e:threshold_rule} to $j$ in the neighborhood of $i$.
In addition, for the TD algorithm to be numerically stable --- i.e., dissipate $E$
at every iteration regardless of the choice of $\delta t$ --- the surface tension matrix has
to be \emph{conditionally negative semi-definite}, which implies $\gamma \in
\mathcal S_N$ has to satisfy
\footnote{Condition \eqref{e:stability_condition} implies $\gamma$ is a matrix that is negative semi-definite as a quadratic form
    on $(1,\dots,1)^\perp$.}
\begin{equation}
 \sum^N_{i,j} \gamma_{ij}\xi_i \xi_j \leq 0 \; 
  \;
 \mathrm{whenever\; } \sum^N_{i}\xi_i=0. 
 \label{e:stability_condition}
\end{equation}
If condition~\eref{e:stability_condition} fails, the algorithm
could lead to erroneous movement of a grain boundary network;  
possibly, the TD algorithm causes the local energy of the system to increase, 
despite the overall free energy decreasing.
While \citet{Esedoglu:2015} proved that the stability
condition~\eref{e:stability_condition} is satisfied for energies sampled from
the RSE form regardless of the size $N$ of the system, the stability condition may not hold for general GB energy
functions. 

In addition, Algorithm~\ref{algo:salvador} also requires the matrix of
reciprocal mobilities $1/\mu$, with entries $\mu_{ij}^{-1}$, to be conditionally negative
semi-definite.
If $\gamma$  and $1/\mu$ are conditionally negative-semidefinite, 
a judicious choice of parameters $\alpha$ and $\beta$ will ensure
the unconditional stability of the algorithm. 
In Ref.~\citep{Salvador:2019}, it was shown that
\begin{equation}
    \alpha \geq \frac{ \displaystyle{
    \min_{i=1,...,N-1 } s_i }}{\displaystyle{
    \max_{i=1,...,N-1 } m_i }} \quad
    \text{and} \quad
    \beta \leq \frac{ \displaystyle{
    \max_{i=1,...,N-1 } s_i }}{\displaystyle{
    \min_{i=1,...,N-1 } m_i }}
    \label{e:stability_choice_2}
\end{equation}
guarantees unconditional stability. 
In \eqref{e:stability_choice_2}, $s_i$ and
$m_i$ are the nonzero eigenvalues of matrices $J\gamma J$ and $J
\frac{1}{\mu}J$, where $J=I-\frac{1}{N} \bm e \otimes  \bm e$ and $\bm
e=(1,\dots,1)$. Note that $\alpha$ and $\beta$ that satisfy \eqref{e:stability_choice_2} automatically guarantee the positiveness of the kernel ($K>0$), which is an essential condition for TD algorithms to attain the viscosity solution for corresponding interfacial motion~\citep{Threshold:viscosity}.
%~\footnote{A viscosity solution satisfies the given partial differential equation almost everywhere and also satisfies certain inequalities expressed with respect to test functions (sub-solutions and super-solutions) at points where the solution may not be differentiable.
%In the present context, it means unique evolution families of surfaces~\citep{ViscositySolution:Goto}. }

\begin{algorithm}
\caption{Threshold-dynamics of \citet{Salvador:2019}}
\label{algo:salvador}
\begin{algorithmic}[1]
\renewcommand{\algorithmicrequire}{\textbf{Input:}}
\renewcommand{\algorithmicensure}{\textbf{Output:}}
 \REQUIRE  The surface tension matrix $\gamma$, the mobility matrix $\mu$, the initial grain structure $\Sigma^0_1, ..., \Sigma^0_N $ described on a discrete grid of which size is $\delta x$, and the final time $T$
 \ENSURE $\Sigma^{n+1}_N $ at the $(n+1)$-th time from the grain shapes
    $\Sigma^{n}_1, ..., \Sigma^{n}_N$
    
    \STATE Choose the kernel width $\delta t$ (equivalent to the time step size $t_s$ of the algorithm) ensuring: 
    $$ \delta x < \gamma_{ij}\mu_{ij}\kappa_{\mathrm{min}}\delta t$$
     
    \STATE Choose $\alpha$ and $\beta$ according to \eref{e:stability_choice_2}
    \STATE Construct the two Gaussians $G_{\sqrt{\alpha \delta t}}$ and $G_{\sqrt{\beta \delta t}}$
    \STATE $t=0$
    \WHILE {$t <T$}
    
    \STATE Compute the convolutions:\\
     $$
        \phi^n_{1,i} = G_{\sqrt{\alpha \delta t}}* \mathbb{I}_{\Sigma_i^n} 
    \quad  \mathrm {and} \quad 
        \phi^n_{2,i} = G_{\sqrt{\beta \delta t}}* \mathbb{I}_{\Sigma_i^n}
    $$ 
    
    \STATE Form the comparison functions: 
    $$
        \psi^n_{i}= \sum_{i\neq j}a_{ij}\phi^n_{1,j} + b_{ij}\phi^n_{2,j}
    $$ 
    where $a_{ij}$ and $b_{ij}$ are given by \eref{e:kernel_coeff_a} and \eref{e:kernel_coeff_b}
    
    \STATE Update the grain shapes: 
    
    \begin{equation*}
    \Sigma^{n+1}_i =\{ \bx:\psi^n_i (\bx) < 
    \displaystyle{
    \min_{j\neq i } }\; \psi^n_j (\bx) \}
    \end{equation*}
    
    \STATE $t=t+t_s$
    
    \ENDWHILE

\end{algorithmic}
\end{algorithm}

\section{Simulations}
\label{sec:simulations}

We will now present a statistical study of grain microstructure evolution under anisotropy energies and
mobilities in large two-dimensional polycrystals using
Algorithm~\ref{algo:salvador}. In particular, we focus on the following two
phenomena: 
\begin{enumerate}
    \item Onset of AGG under anisotropic grain boundary energy and mobility
    \item Development of texture during grain growth 
\end{enumerate}

\subsection{Abnormal grain growth}
\label{subsec:abnormal_growth}

AGG refers to the enlargement of a minority of grains
in a polycrystal at the expense of the surrounding grains~\citep{Rollet:2015},
and is widely observed in many systems including thin films.  
AGG contrasts with normal grain growth, characterized by a relatively narrow grain
size range and a self-similar grain size distribution in time~\citep{Atkinson}.
While anisotropy in GB properties has been proposed as the underlying cause of AGG~\citep{Mullins:2002}, quantitatively validating the hypothesis is a challenging task due to the complexity of
GB anisotropy, expressed as a function of five dofs of the grain boundary character. 
Moreover, there is no agreement on which particular microstructural
characteristics signify the existence of AGG~\citep{Rollet:2015}.
This motivates a statistical study on AGG in a simplified system amenable for
further analysis.  

\begin{figure}
    \centering
    \includegraphics[width=0.5\textwidth]{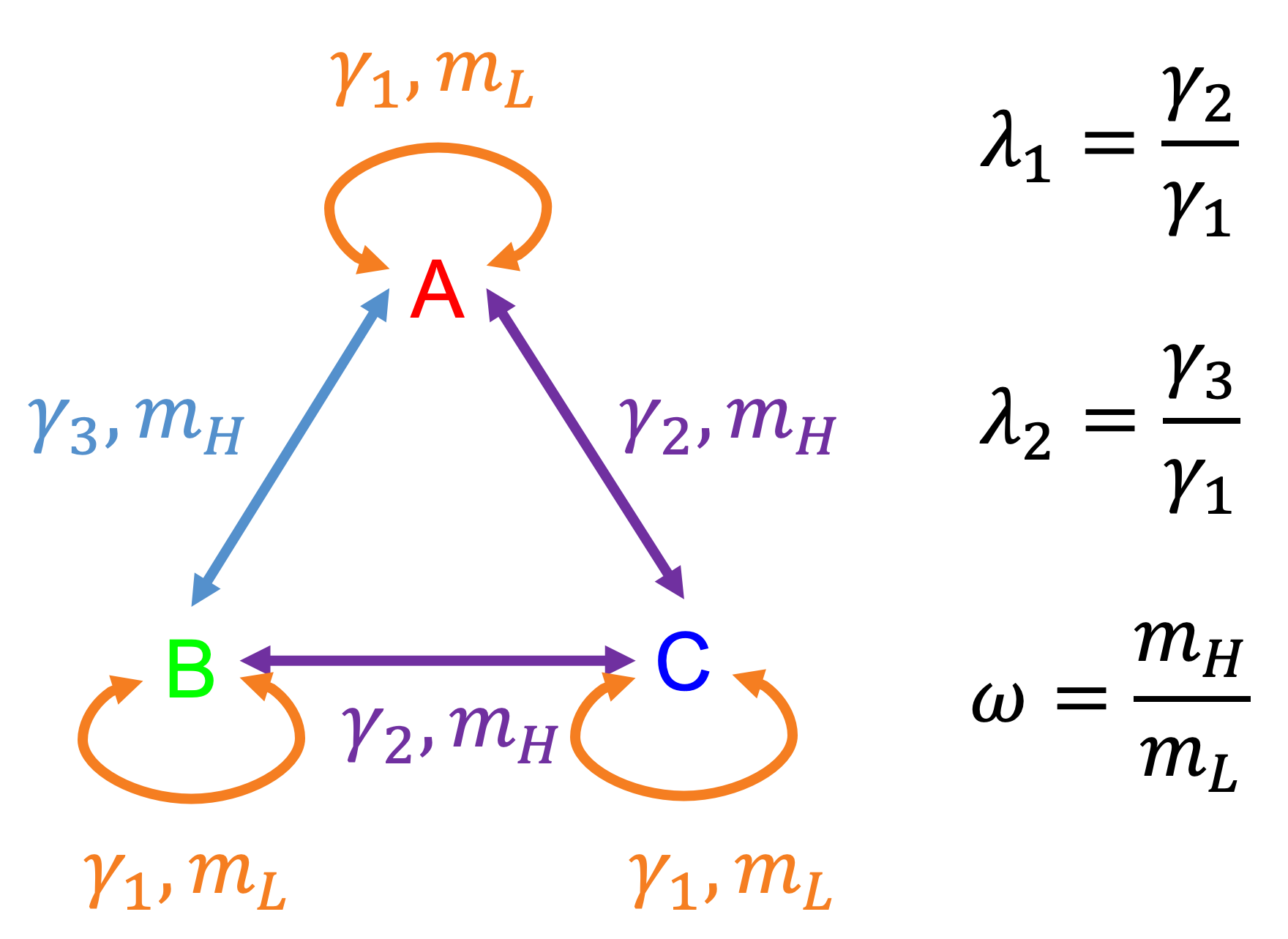}
    \caption{A mnemonic depicting grain boundary anisotropy of a polycrystal used to study
        AGG in \sref{subsec:abnormal_growth}.} 
    \label{fig:three_group_schematic}
\end{figure}

\subsubsection{AGG in an anisotropic tricrystal}
To investigate and analyze AGG, we considered a polycrystal with
grains belonging to three distinct groups (A, B, and C), and described by a few parameters. 
Due to its simplicity, this system serves as a minimal example that leads to AGG. 
We assume that the misorientations between grains within each group are small
and the GBs formed by them have identical energies. On the other
hand, it is assumed that grains belonging to different groups are highly
misoriented and the GB energies between them are larger than the
energy of those formed between grains from the same group. If $\gamma_{ij}$
($i=A, B$ or $C$) denotes the GB energy between the
$i$-th and the $j$-th groups of grains, then we have
$\gamma_{AA}=\gamma_{BB}=\gamma_{CC}:=\gamma_1$ and $\gamma_{AB}$,
$\gamma_{BC}$, and $\gamma_{CA}$ are greater than $\gamma_1$. In addition, we
also assume that $\gamma_{CA}=\gamma_{BC}=:\gamma_2$ is smaller than $\gamma_3:=\gamma_{AB}$.
This implies the orientation angles of grains in group $C$ are closer to
those in $A$ and $B$ than $A$-grains are to $B$-grains.
We also assume the system has only two mobilities --- 
one for small misorientation angle GBs $m_L$ (for the same group) and the other for larger misorientation angles $m_H$ (for different groups). 

The aforementioned GB anisotropy is characterized by three non-dimensional
parameters ---  $\lambda_1 = \gamma_2/\gamma_1$ and
$\lambda_2=\gamma_3/\gamma_1$ describe energy anisotropy, and $\omega = m_H/
m_L$ describes the anisotropy in mobility. 
Note that if all the parameters are equal to one, the system is isotropic.  
The mnemonic in \fref{fig:three_group_schematic} depicts the
GB anisotropy of our system. 
We investigated the following three cases:
\begin{enumerate}
    \item Case 1a: isotropic grain growth, i.e., $\lambda_1=\lambda_2=1.0$ and $\omega=1.0$
    \item Case 1b: energy ratios $\lambda_1=1.5$ and $\lambda_2=2.0$, and mobility ratio $\omega=1.0$
    \item Case 1c: energy ratios $\lambda_1=1.5$ and $\lambda_2=2.0$, and mobility ratio $\omega=1.2$
\end{enumerate}

In Case 1b and 1c, the strengths of GB anisotropy parameters remain \textit{weak}
so that $\gamma$ satisfies the stability condition~\eref{e:stability_condition} of the Algorithm~\ref{algo:salvador}.\footnote{
In Ref.~\cite{Esedoglu:2015}, it is provided that a necessary condition for satisfying \eref{e:stability_condition} is that the matrix $\sqrt{\gamma}$ belongs to $\mathcal T_N$.}
However, in these cases, C-type grains are still energetically favored
and we anticipate unusual growth of C-type growth. 

 \begin{figure}
        \centering
        \begin{subfigure}[b]{0.4\textwidth}
            \centering
        \includegraphics[width=\textwidth]{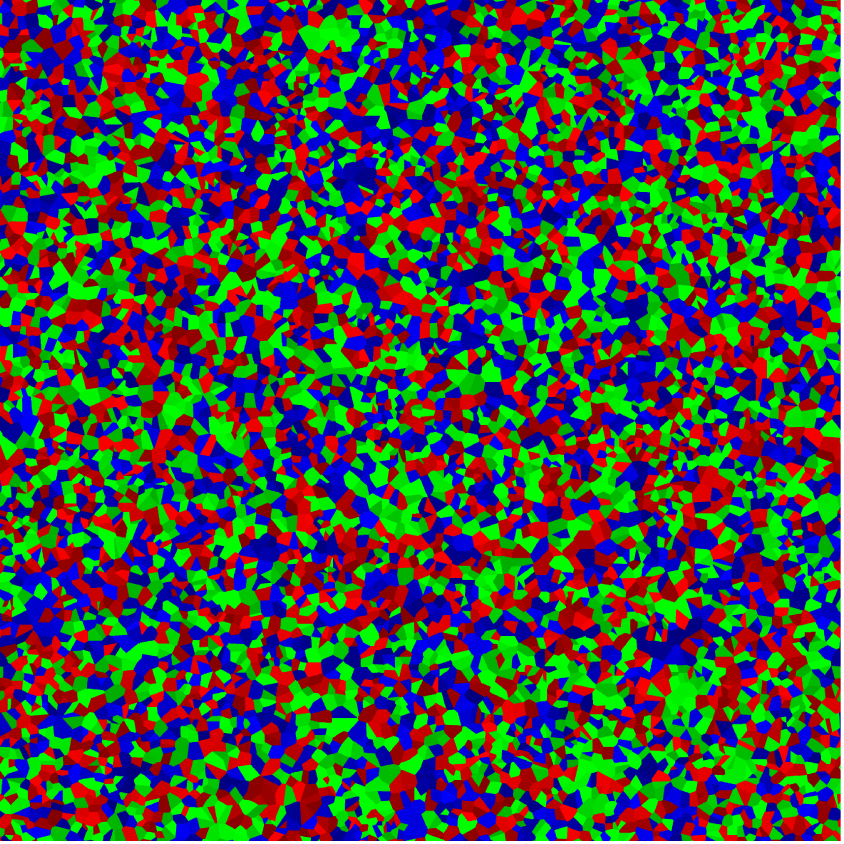}
            \caption{Initial microstructure}
            \label{subfig:ABC_init}
        \end{subfigure}
        \begin{subfigure}[b]{0.4\textwidth}  
            \centering 
            \includegraphics[width=\textwidth]{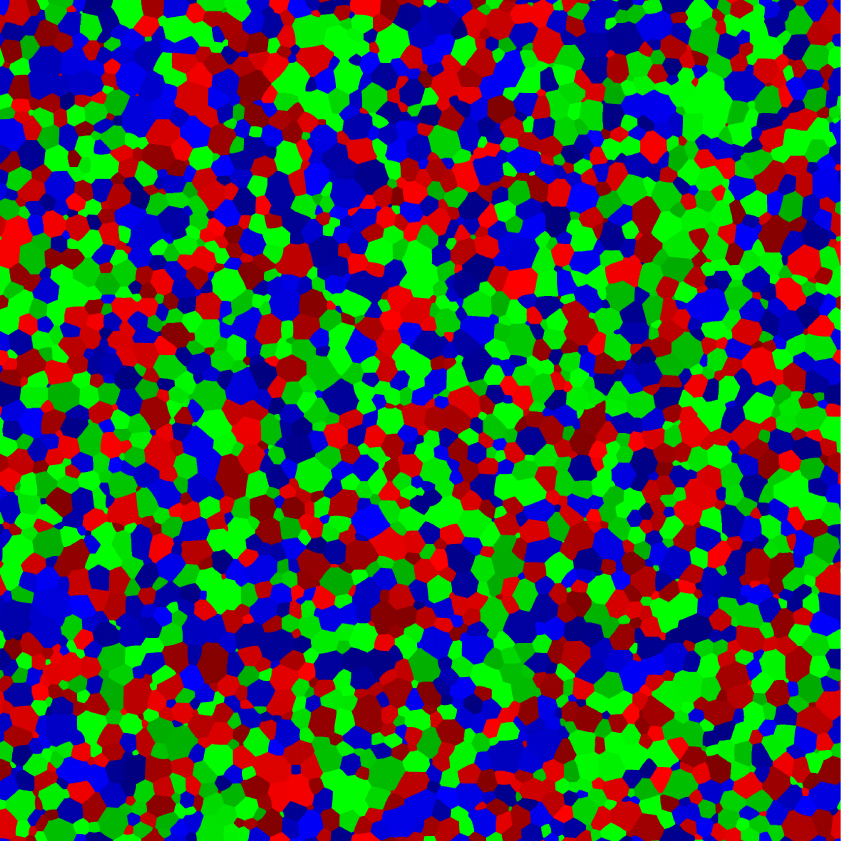}
            \caption{Case 1a}   
            \label{subfig:ABC_Case_1a}
        \end{subfigure}
        \vskip\baselineskip
        \begin{subfigure}[b]{0.4\textwidth}   
            \centering 
            \includegraphics[width=\textwidth]{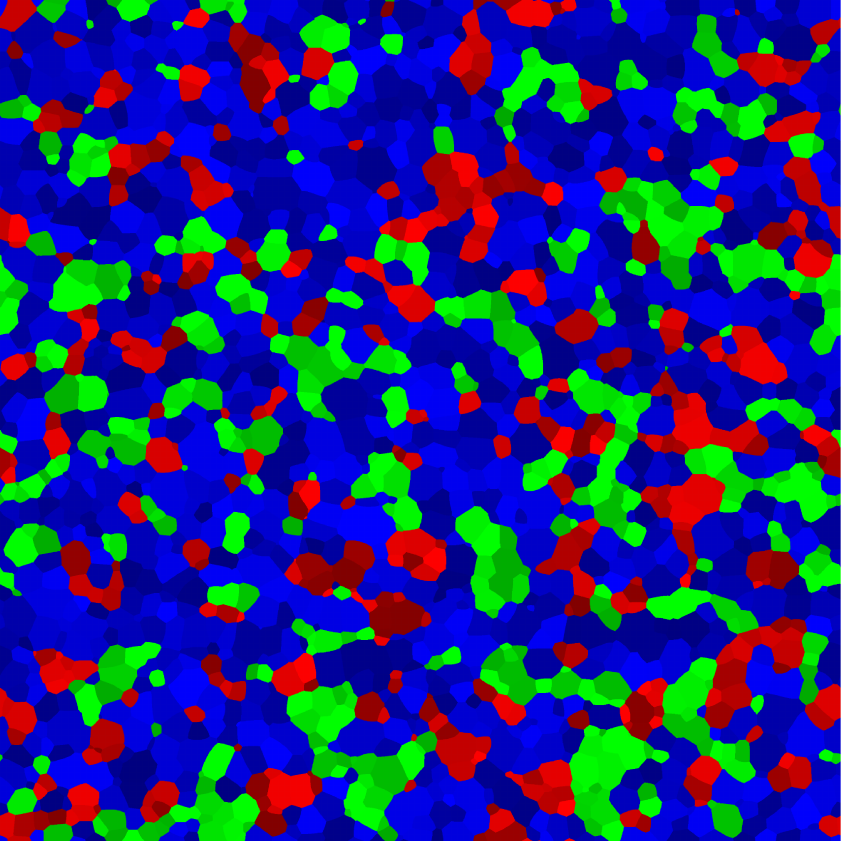}
            \caption{Case 1b}    
            \label{subfig:ABC_Case_1b}
        \end{subfigure}
        \begin{subfigure}[b]{0.4\textwidth}   
            \centering 
            \includegraphics[width=\textwidth]{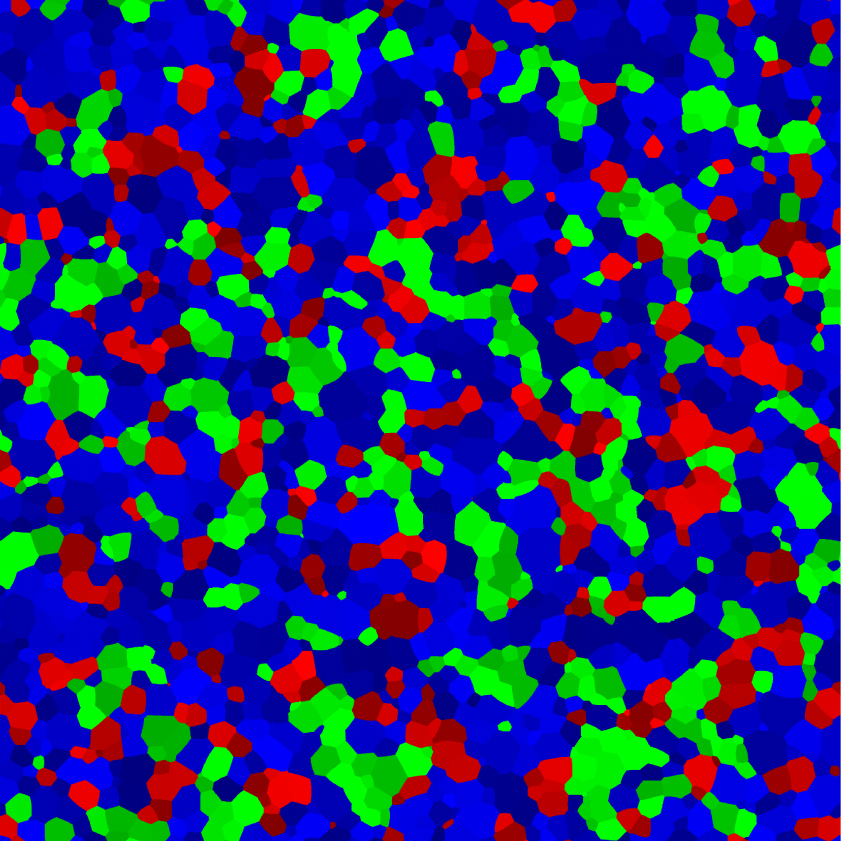}
            \caption{Case 1c}   
            \label{subfig:ABC_Case_1c}
        \end{subfigure}
        \caption{Grain microstructure at time $t_1=500 t_s$
            for the cases summarized in \sref{subsec:abnormal_growth}. 
            The initial microstructure (\protect\subref{subfig:ABC_init}) contains 8,000
            grains. The numbers of remaining grains in 
            (\protect\subref{subfig:ABC_Case_1a}),
            (\protect\subref{subfig:ABC_Case_1b}), and
            (\protect\subref{subfig:ABC_Case_1c}) are 2563, 1984, and 1797, respectively, implying
            GB anisotropy increases the rate of grain coarsening.} 
        \label{fig:microstructure_ABC}
\end{figure}
In all of the cases, we begin with the initial grain microstructure
configuration, shown in \fref{subfig:ABC_init}. 
The initial microstructure consists of a total of $8,000$ grains generated from the
Voronoi tessellations of uniformly distributed random points in the region
$\Omega=[0,1]^2$, discretized by a $3000 \times 3000$ regular grid. Each grain is
randomly assigned to one of three groups from A to C, such that the number
fractions and the area fractions of the three groups are equal to 1/3. 
In \fref{subfig:ABC_init}, the red (group A), green (group B), and blue (group
C) colors distinguish the three groups.

\subsubsection{Results}

Simulations are carried out with a time step size $t_S=5.57\times 10^{-7} \mathsf{t}$
where $\mathsf{t}$ is a unit conversion factor with dimension $[\mathrm{time}][\mathrm{length}]^{-2}$. 
Grain microstructures at time $t_1=2.78\times 10^{-4} \mathsf{t}=500 t_S$ for
the three cases are shown in \fref{subfig:ABC_Case_1a}-\fref{subfig:ABC_Case_1c}.
At the end of the simulations, microstructures of cases 1a, 1b, and 1c contain
2563, 1984, and 1797 grains, respectively. This implies GB anisotropy increases
the rate of grain coarsening. Comparing \fref{subfig:ABC_Case_1a} to
\fref{subfig:ABC_Case_1b} and \fref{subfig:ABC_Case_1c}, we observe that
anisotropy results in clustering of grains belonging to the same groups and an
increase in the area fraction of C-type grains.

\begin{figure}
        \centering
        \begin{subfigure}[b]{0.43\textwidth}
            \centering
        \includegraphics[width=\textwidth]{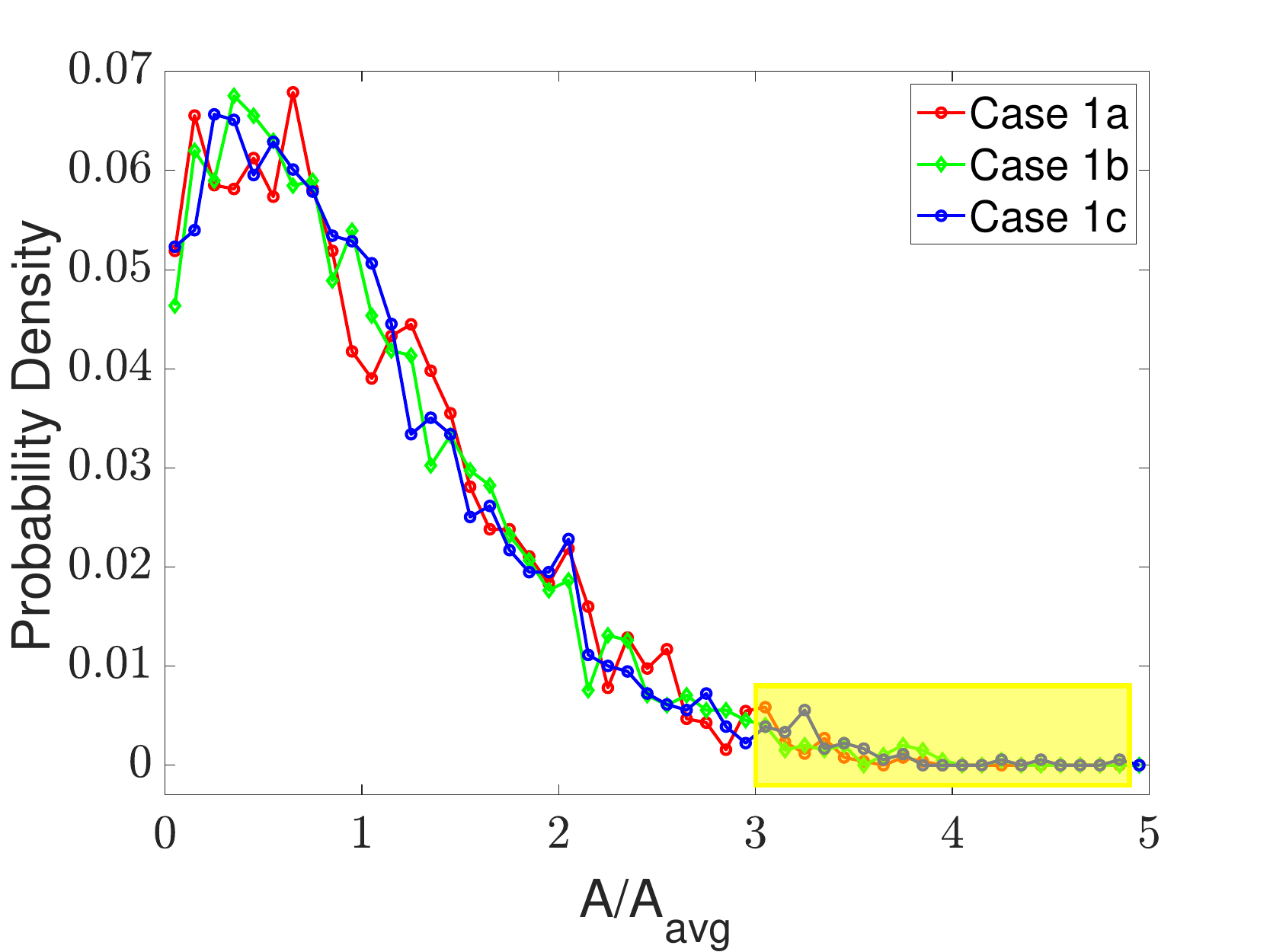}
            \caption{Grain size distribution}
            \label{subfig:GrainSizedistribution_ABC}
        \end{subfigure}
        \begin{subfigure}[b]{0.43\textwidth}  
            \centering 
            \includegraphics[width=\textwidth]{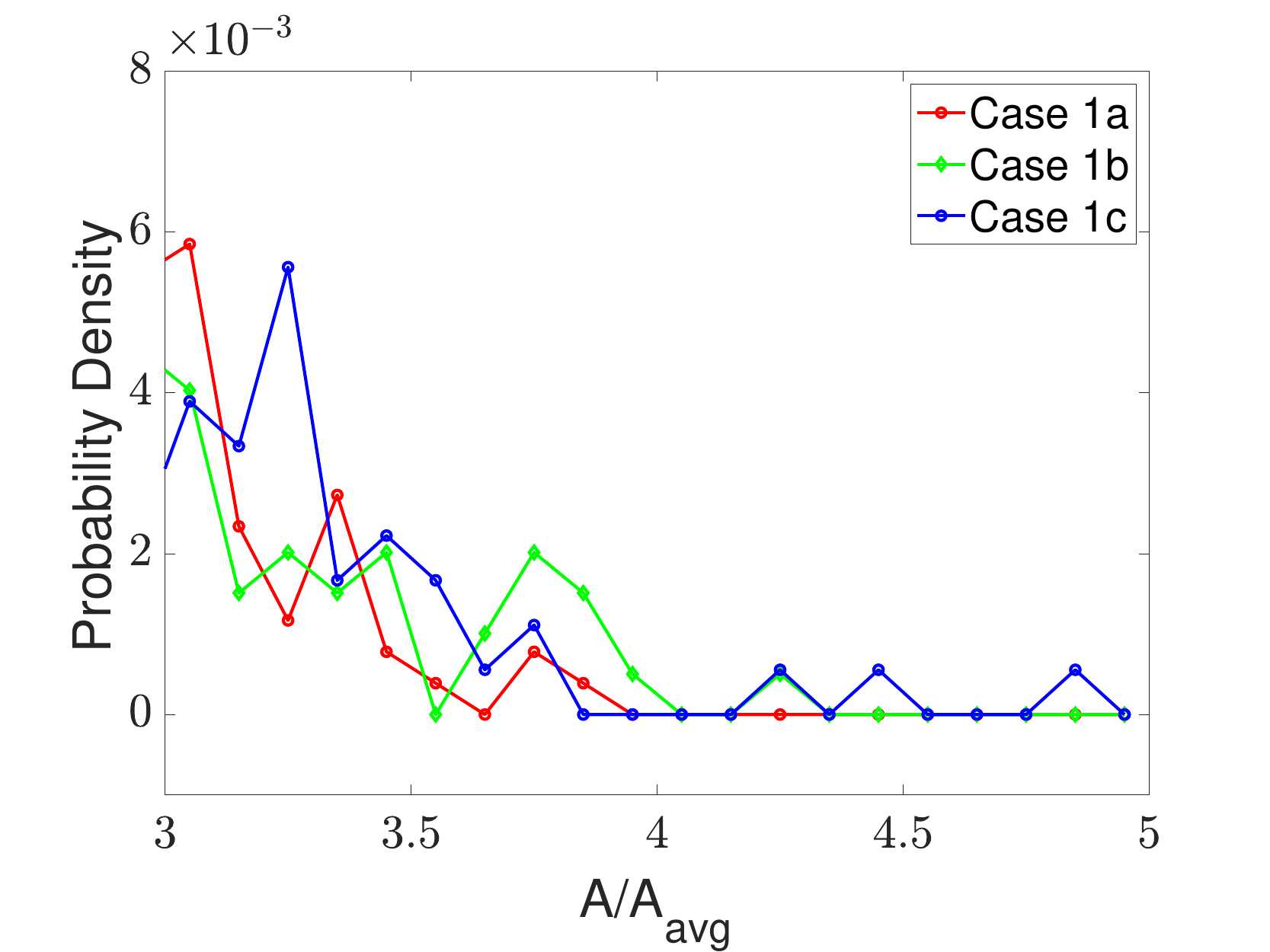}
            \caption{Grain size distribution at tale}   
            \label{subfig:GrainSizedistribution_ABC_zoom}
        \end{subfigure}
        \caption{Grain size distributions of the cases considered in \sref{subsec:abnormal_growth}. A closeup of the tail region, depicted in yellow shade in (\protect\subref{subfig:GrainSizedistribution_ABC}), shows that abnormally larger grains ($A/A_{\text{avg}}\geq 3.5$) are only observed in Case 1b and Case 1c. }
        \label{fig:GrainSizedistribution_ABC}
\end{figure}
Next, we examine and compare the statistical features of the microstructure evolution in the three cases. \fref{subfig:GrainSizedistribution_ABC} shows
plots of grain size distributions at $t=t_1$. The $x$-axes represent
grain areas normalized by the average grain area at $t=t_1$, and the $y$-axes 
represents the probability density, which implies the areas under the graphs are
equal to one. To compute the probability density, a bin size of $0.1$ was used
to partition the $x$-axes. From \fref{subfig:GrainSizedistribution_ABC}, we
conclude that the three grain size distributions are similar for most of the normalized grain areas. 
However, comparing the distributions is a delicate exercise as
abnormally grown grains are spatially
rare~\citep{Rollet:2015}, and therefore, AGG manifests in the tail of a
distribution. A closeup of the tails (yellow shaded region in
\fref{subfig:GrainSizedistribution_ABC}) of the distributions is shown in 
\fref{subfig:GrainSizedistribution_ABC_zoom}, which clearly shows that  abnormally larger grains
($A/A_{\mathrm{avg}}\geq 3.5$) are observed only in cases 1b and 1c. 

\begin{figure}
        \centering
        \begin{subfigure}[b]{0.32\textwidth}
            \centering
        \includegraphics[width=\textwidth]{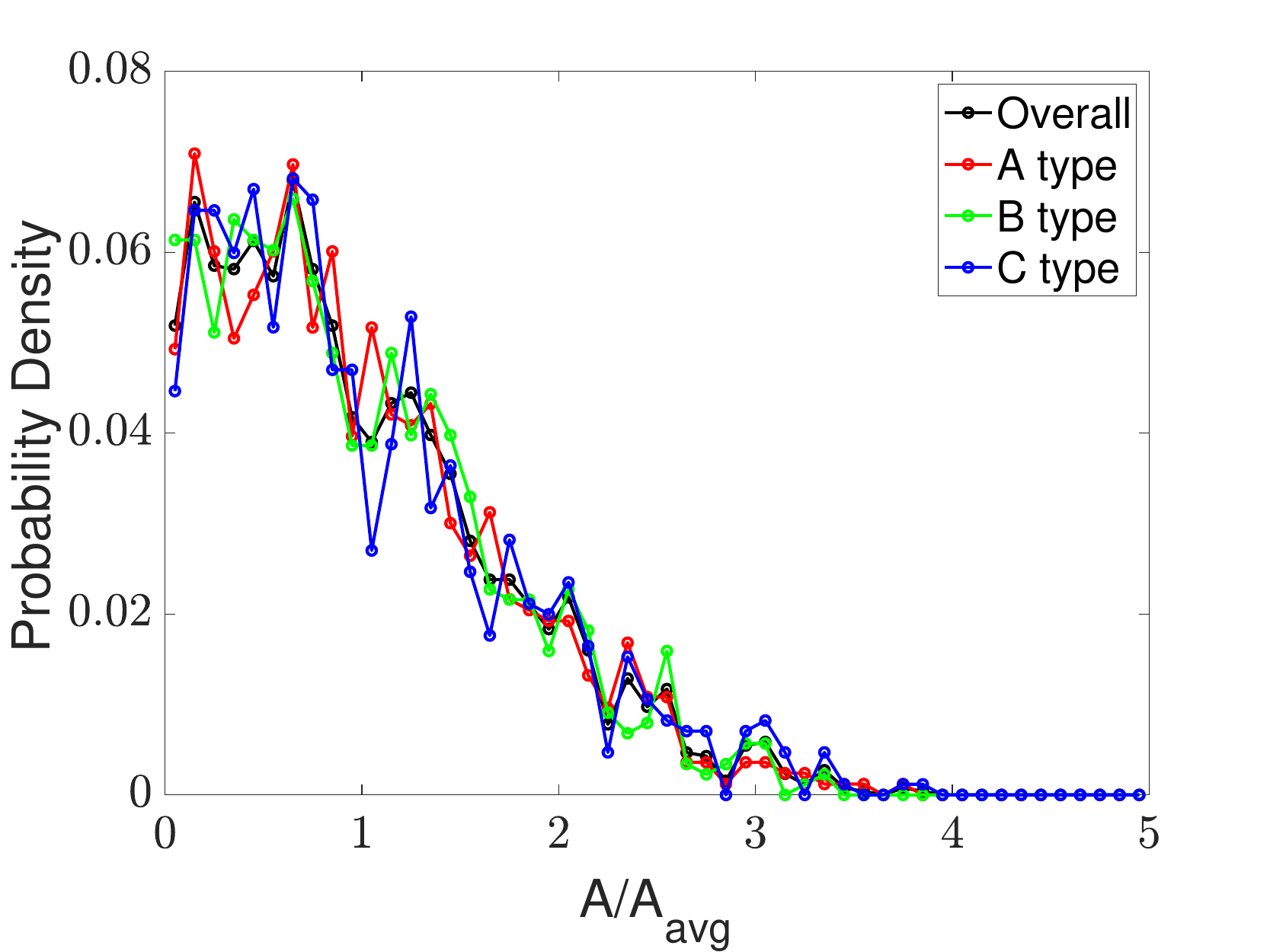}
            \caption{Case 1a}
            \label{subfig:GrainSizedistribution_ABC_separate_a}
        \end{subfigure}
        \begin{subfigure}[b]{0.32\textwidth}  
            \centering 
            \includegraphics[width=\textwidth]{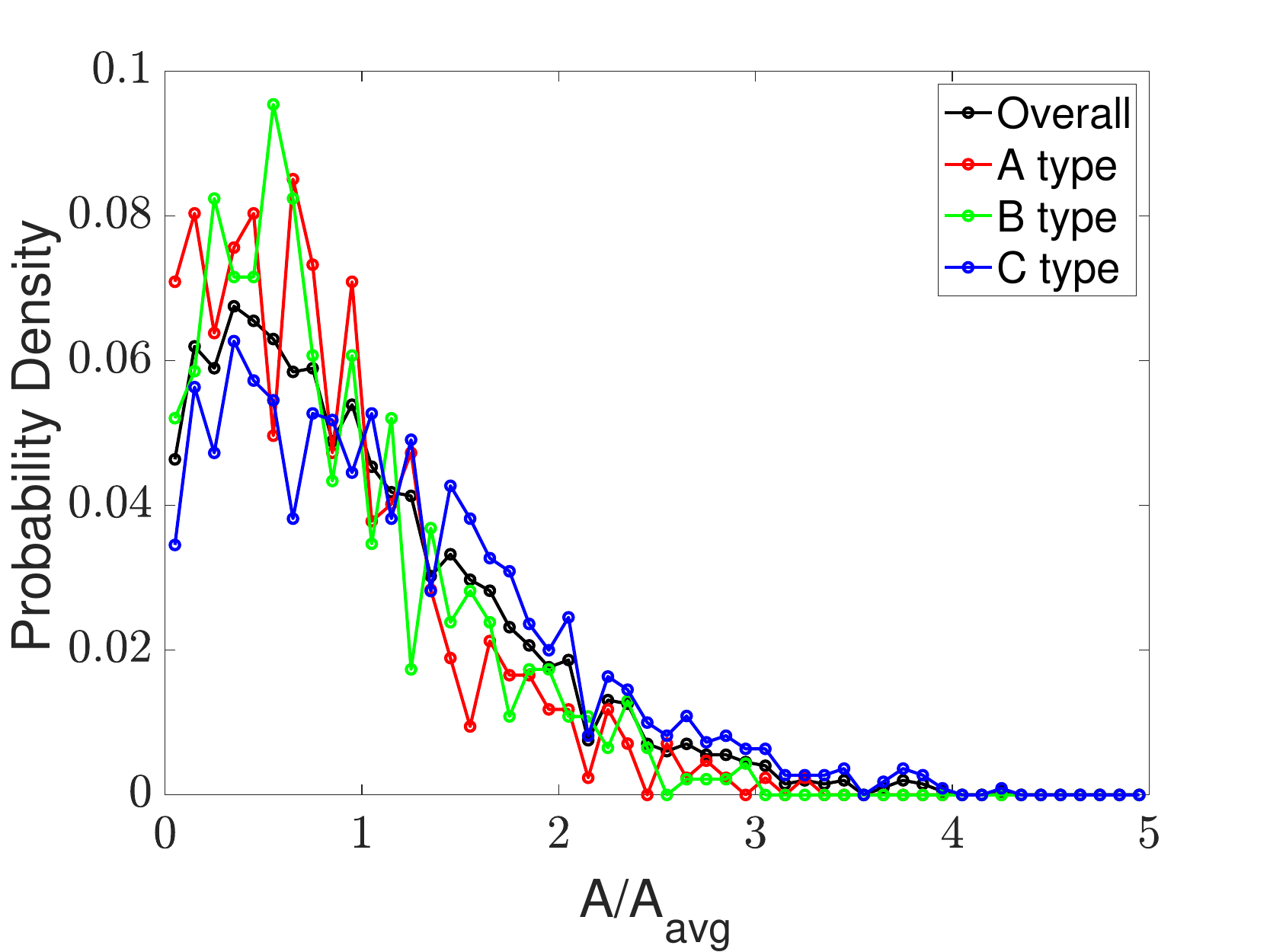}
            \caption{Case 1b}   
            \label{subfig:GrainSizedistribution_ABC_separate_b}
        \end{subfigure}
        \begin{subfigure}[b]{0.32\textwidth}  
            \centering 
            \includegraphics[width=\textwidth]{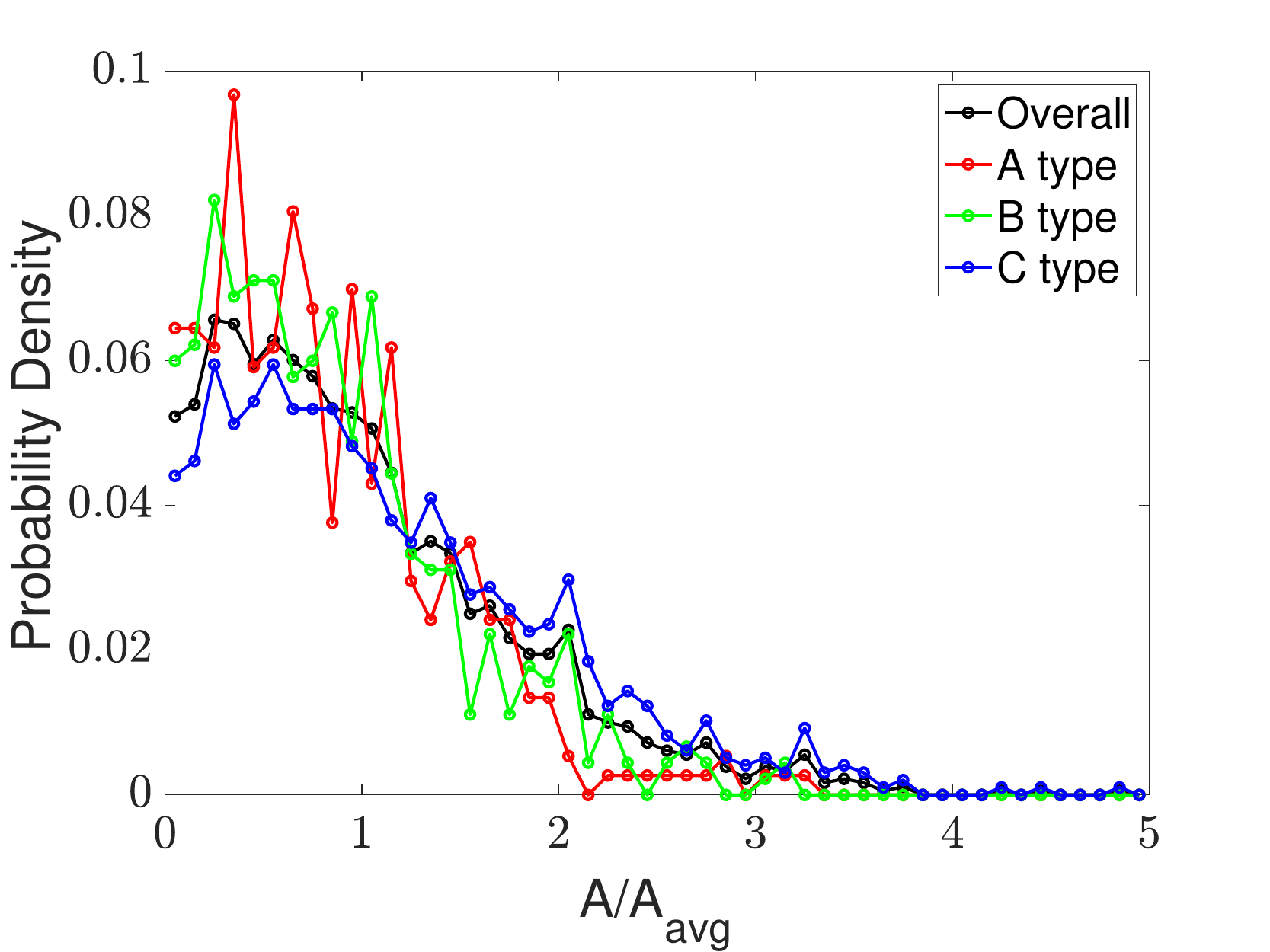}
            \caption{Case 1c}   
            \label{subfig:GrainSizedistribution_ABC_separate_c}
        \end{subfigure}
        \caption{Grain size distributions of each group of grains for the three cases introduced in \sref{subsec:abnormal_growth}. The results of anisotropic case (\protect\subref{subfig:GrainSizedistribution_ABC_separate_b}, \protect\subref{subfig:GrainSizedistribution_ABC_separate_c}) show that the C groups are statistically preferred in Case 1b and Case 1c, resulting in a distinguished grain size distribution from the overall size distribution.   }
        \label{fig:GrainSizedistribution_ABC_separate}
\end{figure}
To further investigate the role of each grain type, 
grain size distribution of each group is separately plotted in \fref{fig:GrainSizedistribution_ABC_separate}. 
Similar to \fref{fig:GrainSizedistribution_ABC}, $x$-axes represents normalized
grain sizes. However, the $y$-axes represent densities (number fractions)
calculated for individual groups. While the distributions of grain types are
identical in the isotropic case
(\fref{subfig:GrainSizedistribution_ABC_separate_a}), the C-type grain has an
extended tail and a smaller peak value in the presence of anisotropy in cases 1b
and 1c (\fref{subfig:GrainSizedistribution_ABC_separate_b} and
\fref{subfig:GrainSizedistribution_ABC_separate_c}). In other words, anisotropy
results in the abnormal growth of C-type grains and lower (relative to the
isotropic case) number of smaller grains. 
The latter is not only a consequence
of AGG but also grain shrinkage, which can be reasoned as follows. If a
C-type grain is surrounded by grains of similar type but not C-type, it is energetically favorable
for it to shrink as $\gamma_1<\gamma_2$. On the other hand, if a C-type grain is
surrounded by grains of different types (A or B), growth is preferred since
$\gamma_2<\gamma_3$.

\begin{table}[t]
\begin{center}
\begin{tabular}{|c|c|c|c|c|}
\hline
                         &                 & A      & B      & C      \\ \hline
\multirow{2}{*}{Case 1a} & Number Fraction & 0.3246 & 0.3433 & 0.3320 \\ \cline{2-5} 
                         & Area Sum   & 0.3236 & 0.3383 & 0.3381 \\ \hline
\multirow{2}{*}{Case 1b} & Number Fraction & 0.2070 & 0.2504 & 0.5426 \\ \cline{2-5} 
                         & Area Sum & 0.1730 & 0.2143 & 0.6127 \\ \hline
\multirow{2}{*}{Case 1c} & Number Fraction & 0.2132 & 0.2324 & 0.5544 \\ \cline{2-5} 
                         & Area Sum   & 0.1739 & 0.1957 & 0.6304 \\ \hline
\end{tabular}
\caption{The number and area fractions of each group at time $t_1=500t_s$ for Case 1a to 1c.}
\label{tab:ABC_fraction}
\end{center}
\end{table}

Table~\ref{tab:ABC_fraction} lists the
number fraction and areas of grains of different types measured in the three
case studies. The number fraction and the area of C-type grains is the maximum
in Case 1c. This quantitative comparison shows that mobility anisotropy (Case
1c) enhances the effects of energy anisotropy (Case 1b).

\begin{figure}
    \centering
    \begin{subfigure}[b]{0.32\textwidth}
        \centering
    \includegraphics[width=\textwidth]{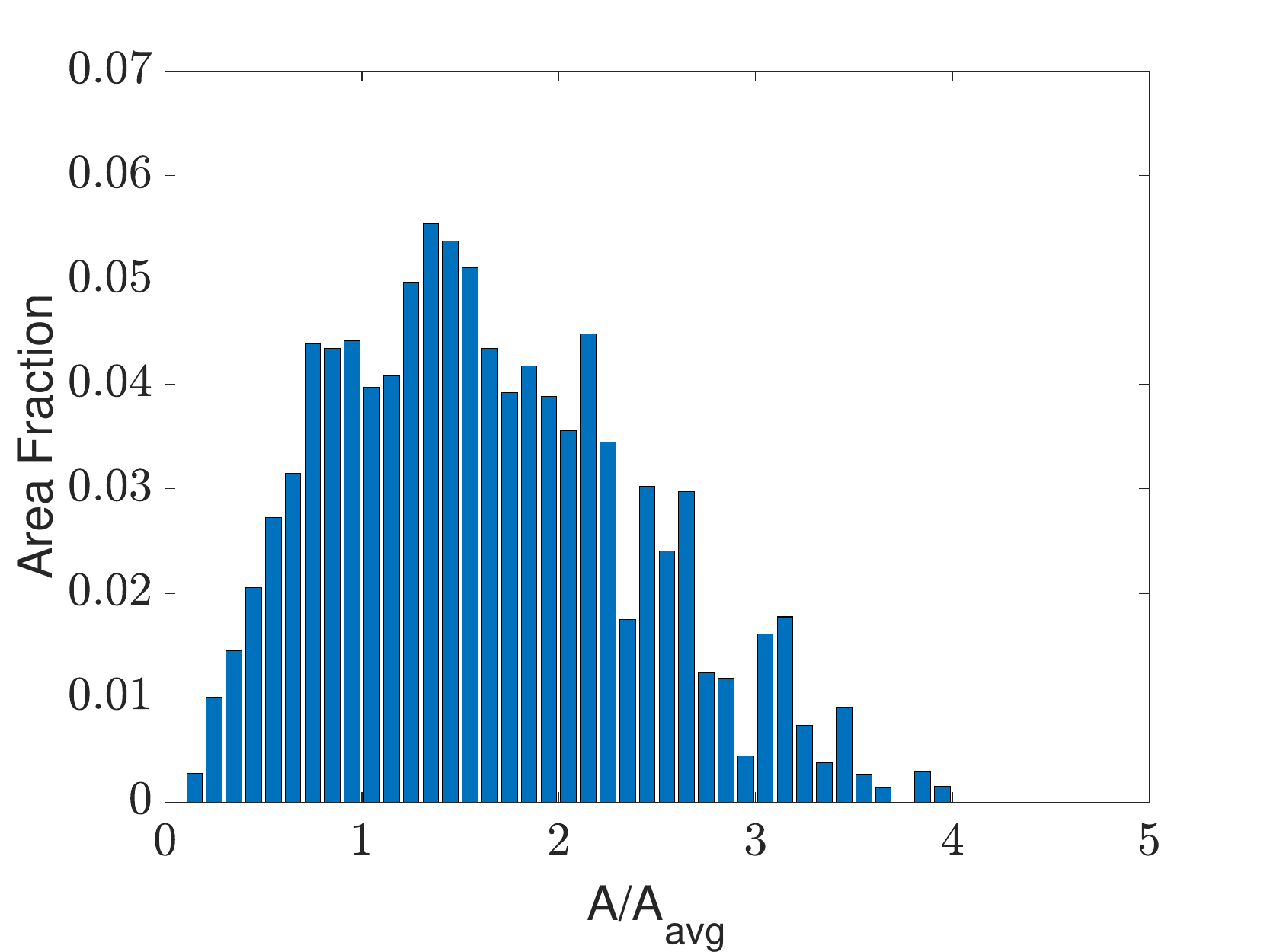}
        \caption{Case 1a}
    \end{subfigure}
    \begin{subfigure}[b]{0.32\textwidth}  
        \centering 
        \includegraphics[width=\textwidth]{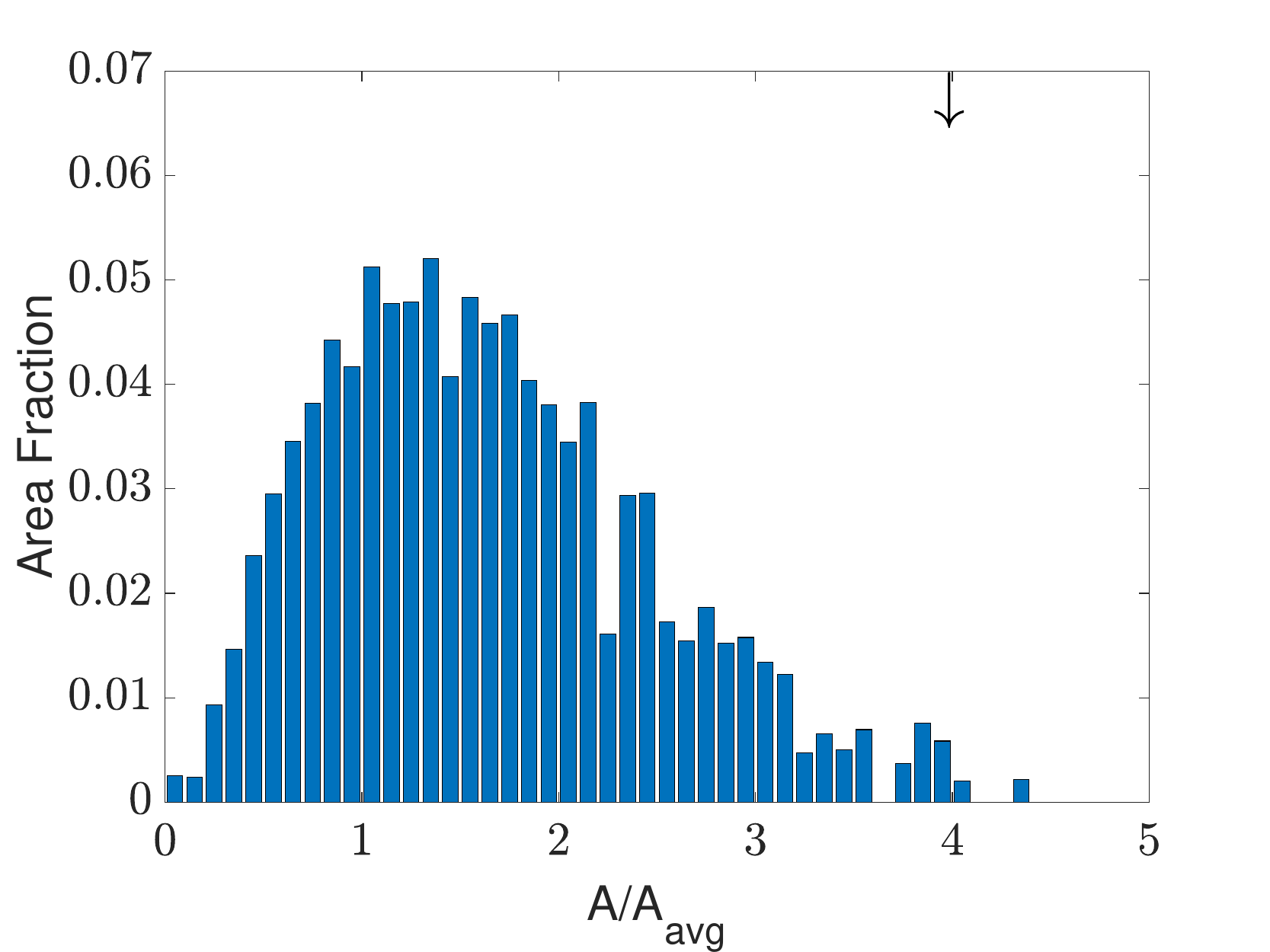}
        \caption{Case 1b}   
        \label{subfig:GrainSizedistribution_ABC_areafraction_b}
    \end{subfigure}
    \begin{subfigure}[b]{0.32\textwidth}  
        \centering 
        \includegraphics[width=\textwidth]{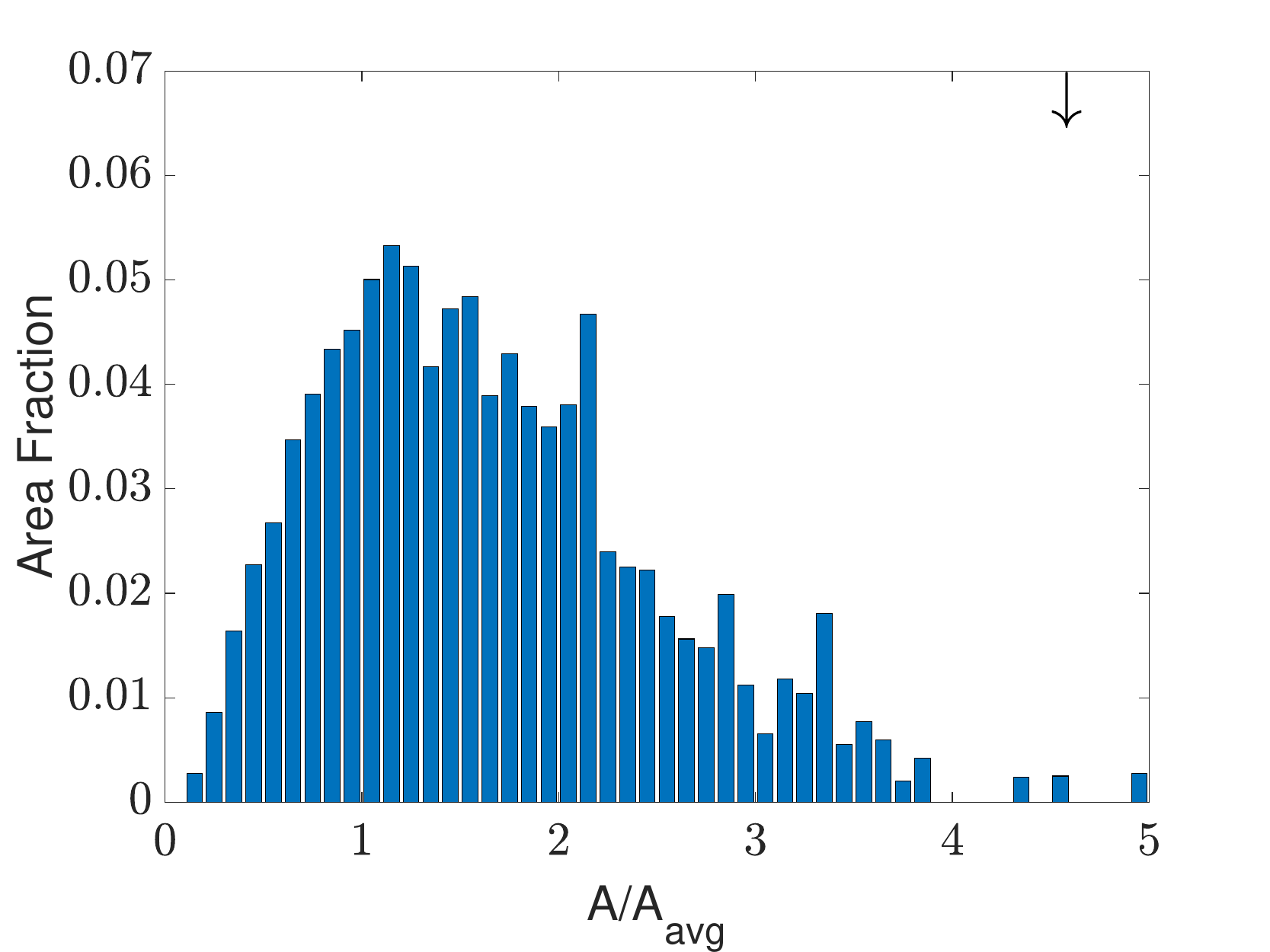}
        \caption{Case 1c}   
        \label{subfig:GrainSizedistribution_ABC_areafraction_c}
    \end{subfigure}
    \caption{Grain size distribution in terms of area fractions. The same bins are used as \fref{fig:GrainSizedistribution_ABC}.  
    The grain boundary energy anisotropy (\protect\subref{subfig:GrainSizedistribution_ABC_areafraction_b}) triggers a bimodular distribution, 
    while anisotropy mobility (\protect\subref{subfig:GrainSizedistribution_ABC_areafraction_c}) moves
    the location of the second mode further right.}
    \label{fig:GrainSizedistribution_ABC_areafraction}
\end{figure}

\fref{fig:GrainSizedistribution_ABC_areafraction} shows plots of \textit{area
fractions} versus the normalized grain sizes for the three cases. Compared to the probability density
plots (\fref{fig:GrainSizedistribution_ABC} and
\fref{fig:GrainSizedistribution_ABC_separate}), AGG is
more conspicuous in the area fraction plots as the smaller number of large-sized grains have a significant effect in the latter plots. In
\fref{subfig:GrainSizedistribution_ABC_areafraction_b}, we observe that GB energy anisotropy
triggers a bimodal distribution, as predicted by a mean-field
theory of \citet{ABBRUZZESE}. In the presence of anisotropic grain boundary
mobilities (Case 1c), \fref{subfig:GrainSizedistribution_ABC_areafraction_c} shows
that the second mode moves further right. 

In summary, GB anisotropy introduces a \textit{statistical preference}
to grains with particular orientations, resulting in 
inhomogeneity in grain size distributions and spatial arrangements
As the strength of anisotropy increases, it will
eventually lead to abnormal grain growth as seen in experiments~\citep{Rollet:2015}. 

\subsection{Texture formation}
\label{subsec:case2}
\begin{figure}
\begin{center}
\includegraphics[width=0.7\textwidth]{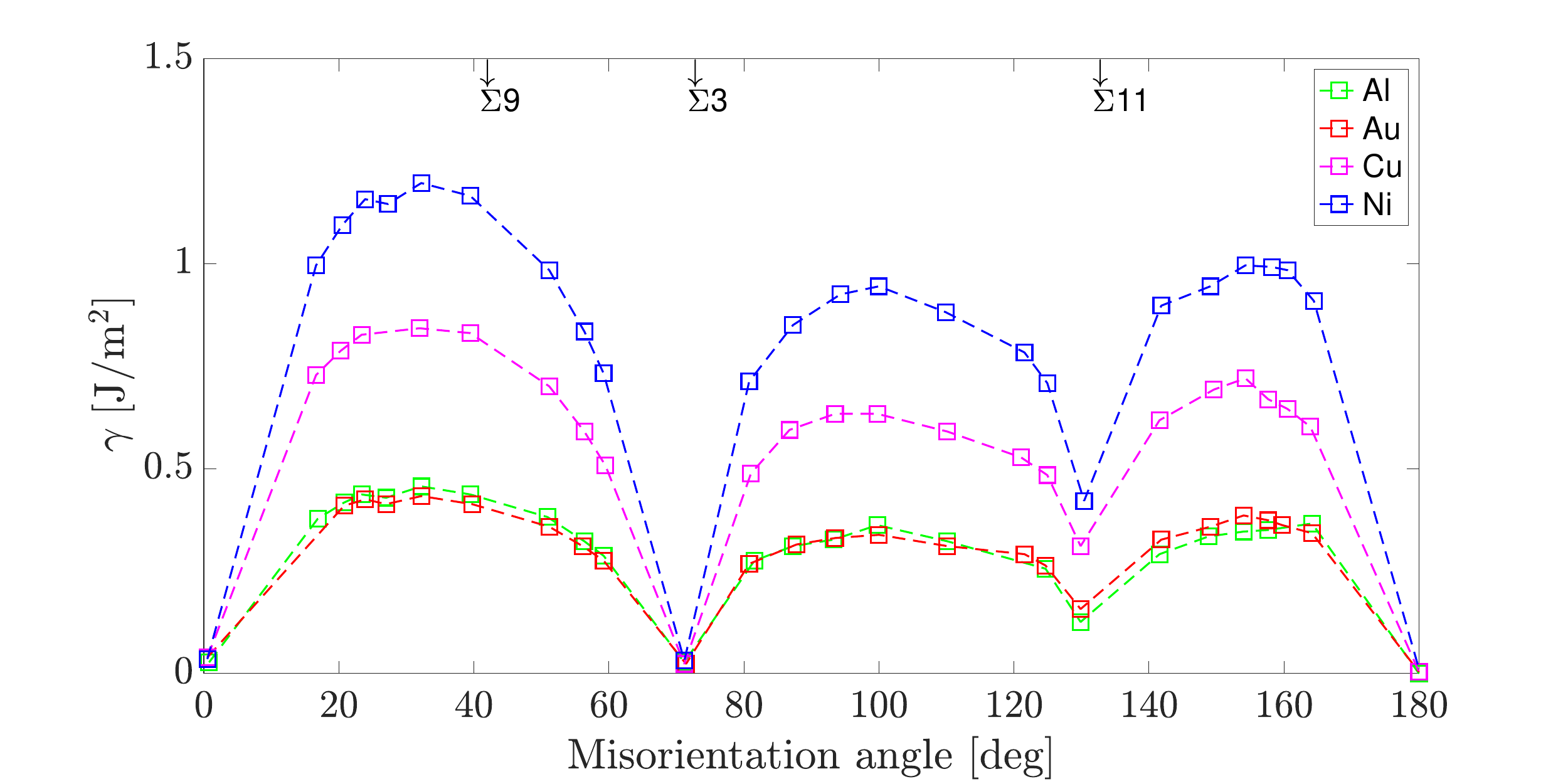}
\end{center}
\caption{Molecular dynamics predictions on grain boundary energy density as a function of misorientation angle for a $[110]$ symmetric-tilt grain boundary in face-centered-cubic materials~\cite{Holm:2010,Bulatov:2014}. Misorientations
corresponding to low energy $\Sigma$ boundaries are marked on the upper axis.}
\label{fig:MDdataset}
\end{figure}

\begin{figure}
\centering
        \begin{subfigure}[b]{0.4\textwidth}
        \centering
        \includegraphics[width=\textwidth]{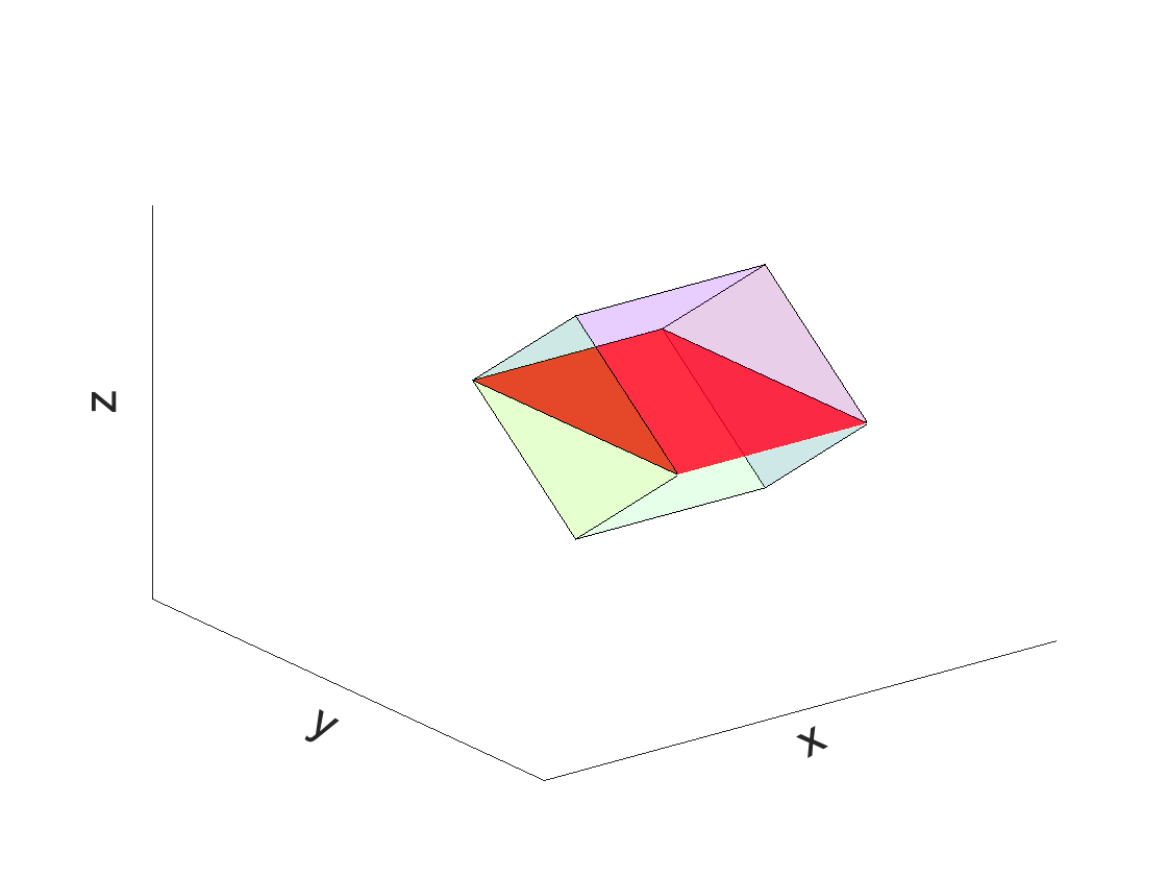}
        \caption{Reference orientation state}
        \label{subfig:110_stgb_energy_ref_state}
        \end{subfigure}
        \begin{subfigure}[b]{0.5\textwidth}  
        \centering 
        \includegraphics[width=\textwidth]{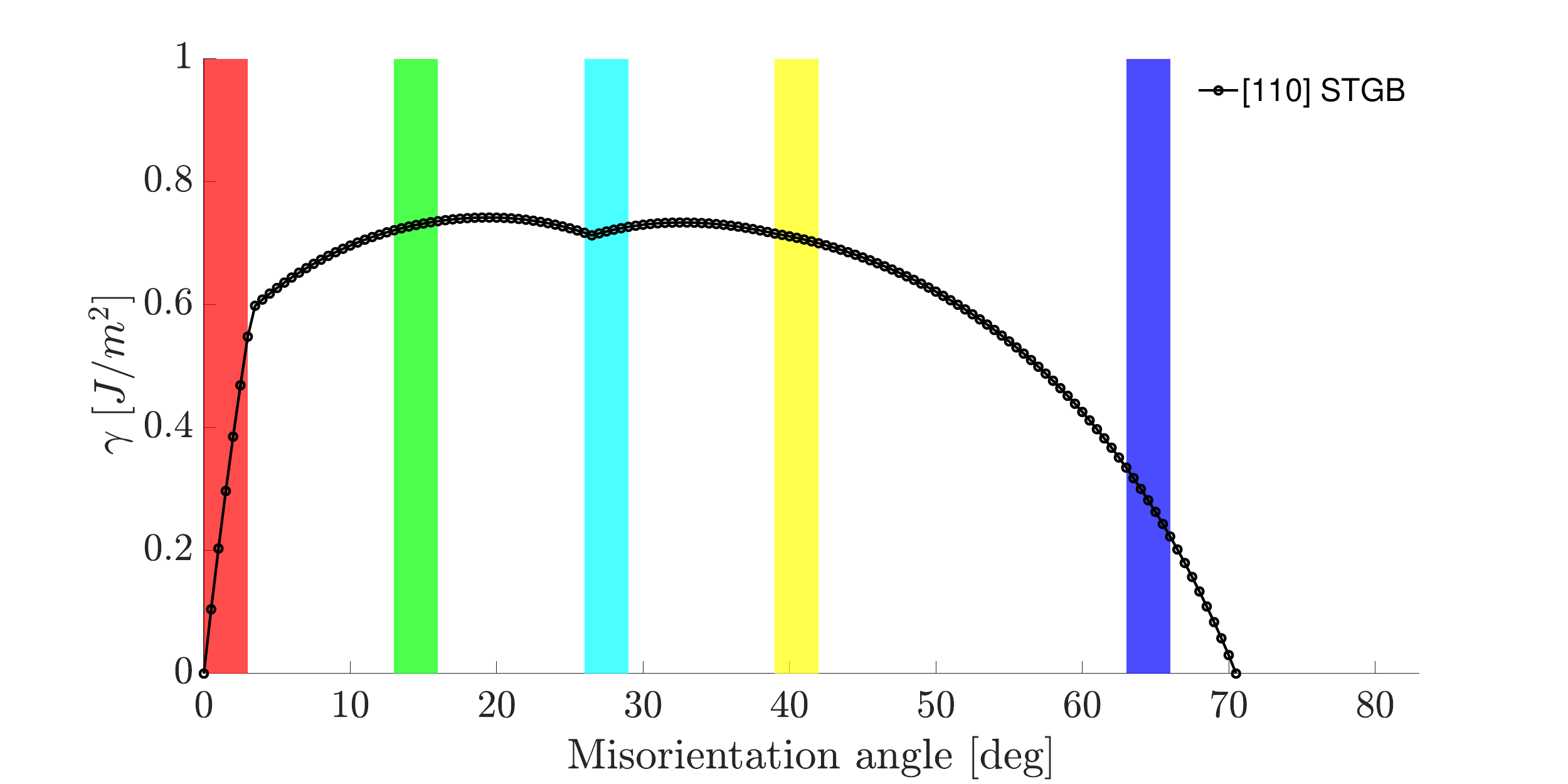}
        \caption{STGB energy}   
        \label{subfig:110_stgb_energy_curve}
        \end{subfigure}
        \caption{
        The [110]-STGB grain boundary energy up to
        misorientation angle $70.6^{\circ}$
	from the reference state shown in (\protect\subref{subfig:110_stgb_energy_ref_state}).
        The sub-domains for group A to E, in which grain orientations are sampled, are colorized in (\protect\subref{subfig:110_stgb_energy_curve}). 
        }        
        \label{fig:110_stgb_energy}      
\end{figure}

In this section, we study the role of grain boundary anisotropy anisotropy and
bicrystallography on the process of texture
formation during grain growth.
For this study, we consider atomistically informed grain boundary energies that
respect the boundary's bicrystallography. For example, \fref{fig:MDdataset}
shows a plot of energy versus misorientation angle for a [110] symmetric tilt
grain boundary energy (STGB).\footnote{The domain of the plot in
    \fref{fig:MDdataset} is restricted to misorientation angles up to 180$^\circ$ since $\gamma$ is symmetric about 180$^\circ$ misorientation angle due to the lattice
symmetry.} The plot shows characteristic local minima, marked as $\Sigma3$ and
$\Sigma 11$, due to enhanced lattice matching of the
grains~\citep{Runnels:2016_1,Wolf:1990} at certain misorientation angles.
In the following section, we describe the polycrystal and the bicrystallography respecting grain boundary energy function used to explore texture formation. 

\subsubsection{The system}

We consider a two-dimensional  fcc polycrystal with the $\langle 110\rangle$ direction of all grains aligned along the z-axis. All grain boundaries are
assumed to be $[110]$ symmetric-tilt type. The orientations of the grains are measured
with respect to a reference grain whose $[100]$ direction is aligned along the
$x$-axis, as depicted in \fref{subfig:110_stgb_energy_ref_state}.
Under these constraints, we only need a single scalar $\theta_i$ 
to describe the orientations of grain samples. 

To strategically investigate the roles of low and high-angle grain boundaries,  we restricted 
the grain orientations, $\theta_i$, to groups of $3^\circ$-length intervals of the
misorientation angle as follows:
\begin{equation}
    \theta_i \in 
\begin{cases}
\mathrm{Group\; A,\; if \;  } \theta_i \in [0,3]^{\circ}, \\
\mathrm{Group\; B,\; if \; } \theta_i \in [13,16]^{\circ}, \\
\mathrm{Group\; C,\; if \; } \theta_i \in [26,29]^{\circ}, \\
\mathrm{Group\; D,\; if \; } \theta_i \in [39,42]^{\circ}, \\
\mathrm{Group\; E,\; if \; } \theta_i \in [63,66]^{\circ}.
\end{cases}
\label{e:sample_group}
\end{equation}
The above intervals are shown in color in \fref{subfig:110_stgb_energy_curve}.
Boundaries between grains from the same group are
identified as LAGBs with misorientation  $< 3^{\circ}$. 
HAGBs of the system are formed by grains from different groups, 
and have a misorientation angle $>10^{\circ}$. 
Groups A and E in \eqref{e:sample_group} are constructed such that the misorientation of a grain boundary between an A-type grain and an E-type grain is close to that of a twin
boundary (TB), which has a misorientation angle of 70.6$^\circ$. 
TBs are often desired in grain boundary engineering as they enhance the strength and ductility of a polycrystal~\citep{Rupert:2014,twinProperty}.
The energies of LAGBs are typically lower than those of HAGBs and increase
steeply with misorientation angles (\fref{subfig:110_stgb_energy_curve}). The
energy of a typical HAGB is less sensitive to changes in the misorientation
angle. However, TBs are HAGBs that are exceptions to the above two
properties as can be inferred from \fref{subfig:110_stgb_energy_curve}.

In this study, we explored microstructures \textit{with} and \textit{without} subgrains.  
In the latter, grains are partitioned into subgrains, which
are connected along LAGBs. Subgrains are commonly
observed in polycrystalline materials subjected to plastic deformation followed by
\textit{recovery} at a temperature below the recrystallization temperature. 
The mechanism is as follows. 
Plastic deformation leads to an increase in dislocations, 
which subsequently rearrange during the recovery process to form LAGBs and subgrains. 
Further deformation promotes the rotation of subgrains resulting
in the transformation of LAGBs into HAGBs~\citep{subgrain:1992,subgrain:2005}. 
Recent progress in \textit{severe plastic deformation}, employed during  manufacturing
processes, can be used to facilitate this mechanism~\citep{Langdon,ECAP:2020}.
%This phenomenon is often referred to as \textit{continuous dynamic recrystallization}, 
%which contrasts with {discontinuous recrystallization} that entails
%\textit{nucleation} of new grains near grain boundaries.\anc{I thought we are
%modeling discontinuous recrystallization without the nucleation phase. Do you
%agree?}\jkc{No. If no nucleation, then it is not discontinuous. We are dealing with continuous}

The following initial microstructures are considered in our simulations:
\begin{enumerate}
\item Case 2a: a tricrystal with subgrains, wherein the three primary grains
    belong to A, B, and D groups. TBs are excluded
\item Case 2b: a polycrystal with subgrains. TBs are included, and LAGBs are dominant.
\item Case 2c: a polycrystal without subgrains. TBs are included, and HAGBs are dominant.
\end{enumerate}
In all cases, we examine 5,000 distinct subgrains (or only grains for Case 2c)
arranged in a periodic domain discretized by a regular grid of size
$3000\times3000$. The mobility ratio between high to low angle grain
boundary is again set to $\omega=1.2$.

The initial tricrystal microstructure of Case 2a is shown in \fref{subfig:subgrained_tricrystal_ic}, 
wherein grains and subgrains are colored based on the color keys
(\fref{subfig:110_stgb_energy_curve}) of the groups
they belong to. The tricrystal consists
of A (red), B (green), and D (yellow) type grains. 
The subgrains have the same color as the grain they belong to and are shaded depending on their orientation relative to their parent grain. By construction, boundaries between different colored grains are HAGBs. The microstructure was generated using a Voronoi tessellation of random seeds. 
Orientation groups are determined by the location of Voronoi seeds, 
while specific orientation values are randomly chosen within the domain of each group.
 
Cases 2b and 2c are designed to evaluate the role of TBs. The initial
microstructure for Case 2b is shown in \fref{subfig:subgrained_crystal_ic}, 
wherein the TBs are the boundaries between type A (red) and type E (blue)
grains. It was generated using a two-level Voronoi tessellation~\citep{Bernacki:2022}. 
The coarse-level Voronoi seeds are used to determine group types, while the
$30\times$-refined seeds define specific orientation values. 
This procedure yields a LAGB-dominated polycrystal wherein each grain is comprised of approximately 160 subgrains.
On the other hand, 
The microstructure of Case 2c generated from a single-level Voronoi tessellation
is dominated by HAGB. 
The initial microstructures of cases 2b and 2c, however different, 
have similar uniform grain orientation distributions
(\fref{subfig:texture_LAGB_t000} and \fref{subfig:texture_HAGB_t000}), and therefore, are
non-textured.
\begin{table}[t]
\begin{center}
\begin{tabular}{|c|c|c|c|c|}
\hline
        & Number of subgrains & Number of grains & Grid size   &  Time step $t_S$ \\ \hline
Case 2a & 5000                & 3                & $3000 \times 3000$ & $4.44\times 10^{-6}$  \\ \hline
Case 2b & 5000                & 30              & $3000 \times 3000$ & $2.22\times 10^{-6}$  \\ \hline
Case 2c & -                   & 5000             & $3000 \times 3000$ & $2.22\times 10^{-6}$  \\ \hline
\end{tabular}
\caption{Microstructure and simulation parameters of numerical experiments considered in \sref{subsec:case2}}
\label{tab:case2_parameters}
\end{center}
\end{table}
Table~\ref{tab:case2_parameters} documents the microstructure and simulation parameters of the three case studies of this section.

An implementation of Algorithm~\ref{algo:salvador} for a large-sized system is a delicate exercise. Typically, a moderate quotient $\alpha/\beta$ of the widths of the two Gaussians~\eref{e:algorithm_kernel} is chosen  
so that the two Gaussians, $G_{\sqrt{\alpha \delta t}}$ and $G_{\sqrt{\beta \delta t}}$, 
are well-resolved even at a fairly refined grid~\citep{Salvador:2019}. In a large-scale system with a wide range of grain boundary energies, however, a naive choice of the width parameters~\eref{e:stability_choice_2} easily becomes ill-posed, because the minimum grain boundary energy in the system can be 
arbitrarily small.
To address the issue, we follow the suggestion in Ref.~\citep{Salvador:2019}. 
The idea is to limit the possible minimum misorientation angles between 
any two grains to no more than 0.5$^\circ$ when constructing the two Gaussian kernels.
This is equivalent to setting the minimum energies 
that can be resolved by the algorithm.

\subsubsection{Results}
\begin{figure}
        \centering
        \begin{subfigure}[b]{0.32\textwidth}
            \centering
        \includegraphics[width=\textwidth]{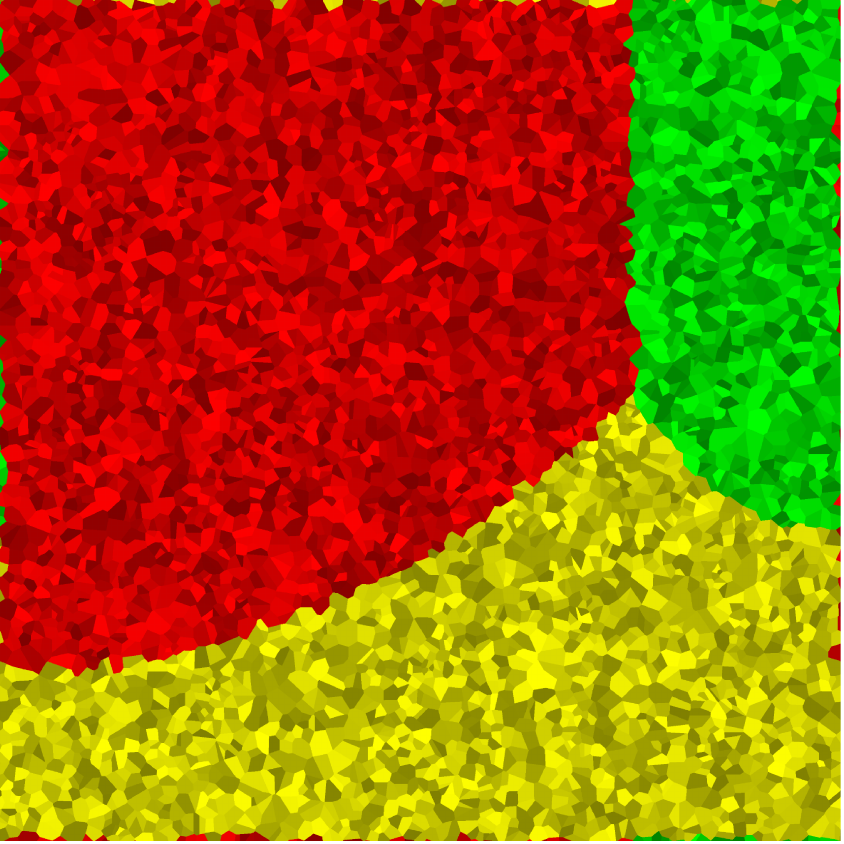}
            \caption{Initial condition}
            \label{subfig:subgrained_tricrystal_ic}
        \end{subfigure}
        \begin{subfigure}[b]{0.32\textwidth}  
            \centering 
            \includegraphics[width=\textwidth]{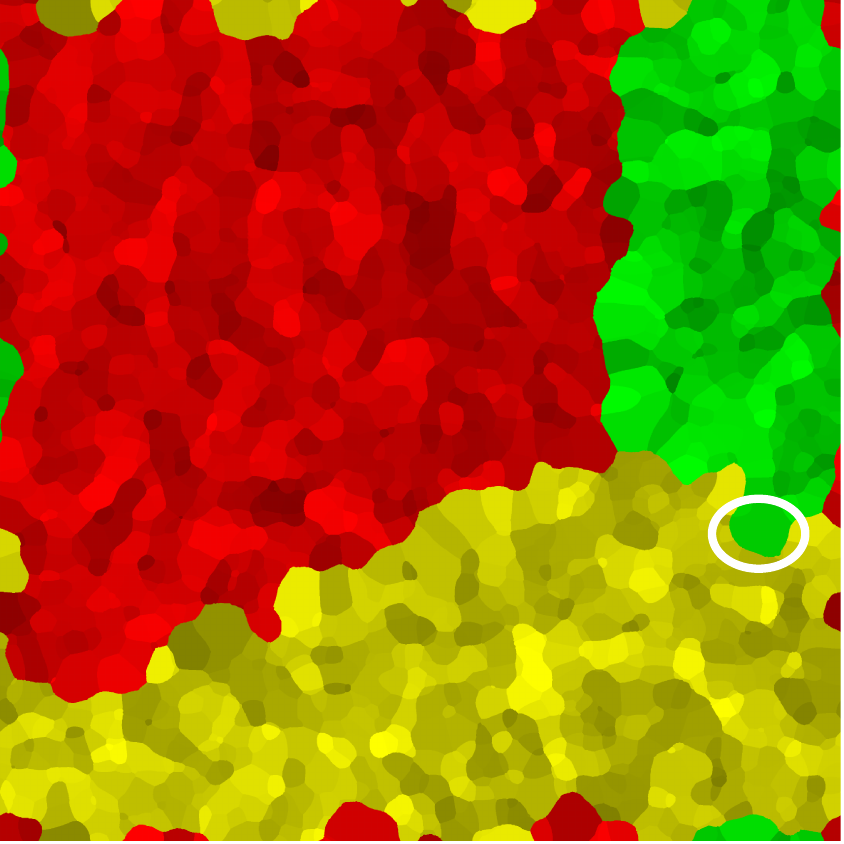}
            \caption{$t=500t_S$}   
            \label{subfig:subgrained_tricrystal_t500}
        \end{subfigure}
        \begin{subfigure}[b]{0.32\textwidth}  
            \centering 
            \includegraphics[width=\textwidth]{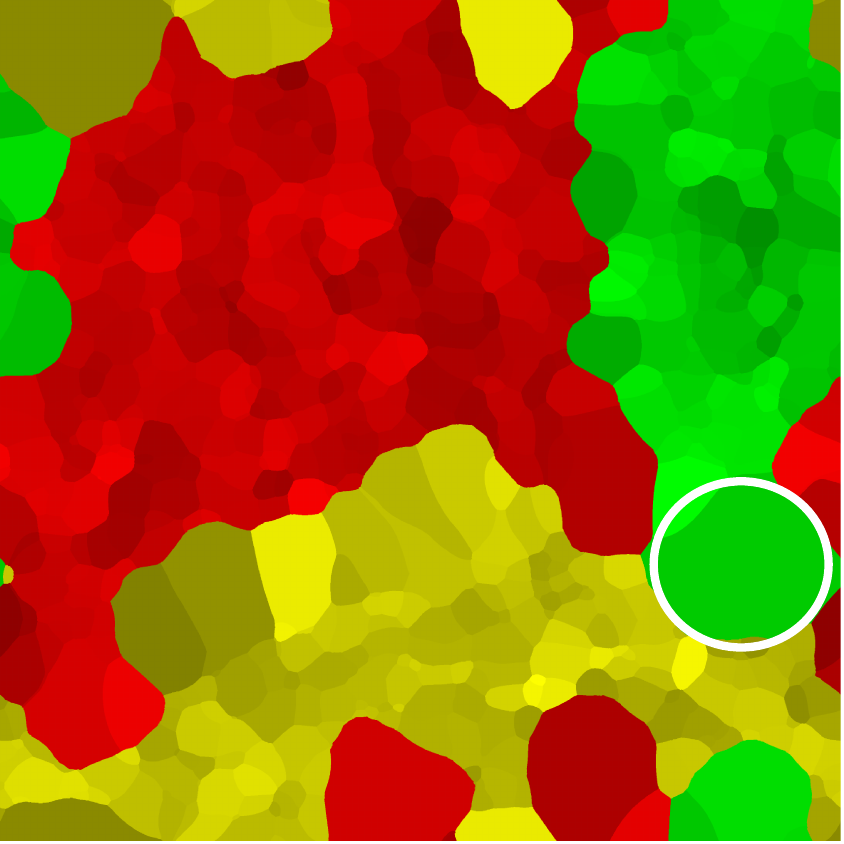}
            \caption{$t=2000t_S$}   
            \label{subfig:subgrained_tricrystal_t2000}
        \end{subfigure}
        \caption{Microstructure evolution of a tricrystal consisting of subgrains (same colored grains). The white circled subgrain is an example of abnormally growing grain near grain boundaries.}
        \label{fig:subgrained_tricrystal}
\end{figure}
\noindent
\underline{\emph{Case 2a}}: 

\fref{subfig:subgrained_tricrystal_ic} through \fref{subfig:subgrained_tricrystal_t2000} show
the time evolution of the tricrystal with subgrains. All subgrains in the initial microstructure have comparable sizes. From
\fref{subfig:subgrained_tricrystal_t500}, we can infer that the grain coarsening
rate is higher near a HAGB because of the higher energy and mobility 
of a HAGB. 
\fref{subfig:subgrained_tricrystal_t500} and \fref{subfig:subgrained_tricrystal_t2000} shows that when a grain (circled) near a HAGB
reaches a critical size, it undergoes AGG. It is important to note that the
critical size plays an important role in selecting grains as not all grains near
HAGBs undergo AGG. A plausible explanation for this observation is that beyond a
critical size, it is energetically favorable for the subgrain to grow in
size and absorb the subgrain boundaries in the adjoining grain. 
The critical size may depend on the relative sizes of the grains compared to neighbors.
As a result, certain subgrains located close to grain boundaries are more
likely to outgrow those located in the interior, as shown in \fref{subfig:subgrained_tricrystal_t2000}. 
It is interesting to note that this mechanism resembles \textit{discontinuous recrystallization}, wherein new defect-free grains nucleate near HAGBs and grow to replace
the microstructure entirely.
\begin{figure}
        \centering
        \begin{subfigure}[b]{0.32\textwidth}
            \centering
        \includegraphics[width=\textwidth]{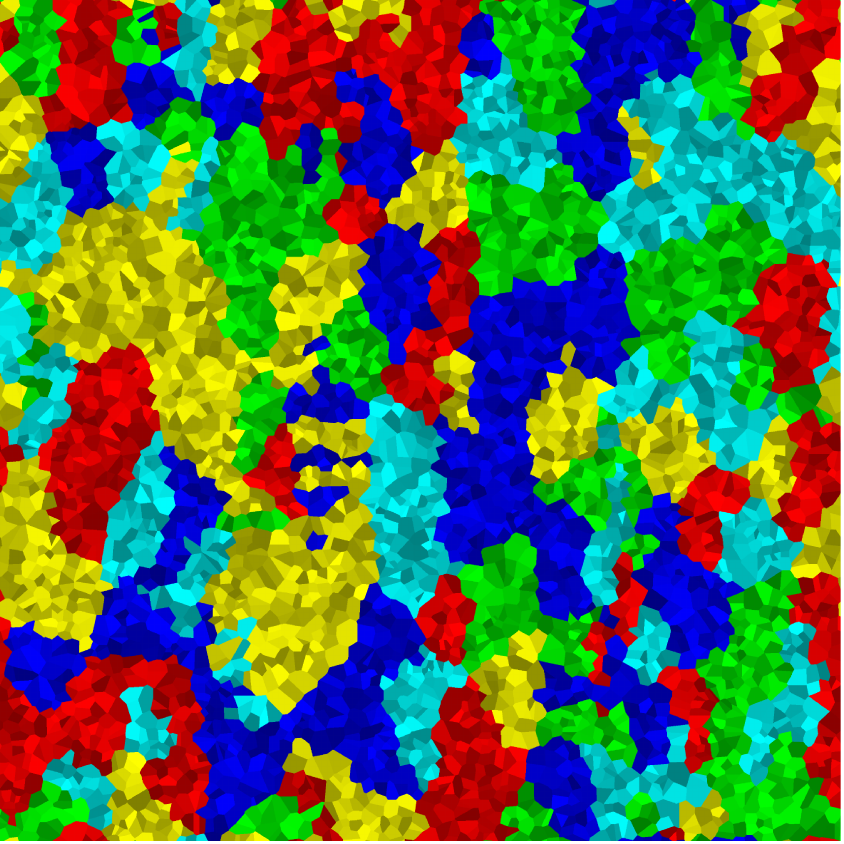}
            \caption{Initial condition}
            \label{subfig:subgrained_crystal_ic}
        \end{subfigure}
        \begin{subfigure}[b]{0.32\textwidth}  
            \centering 
            \includegraphics[width=\textwidth]{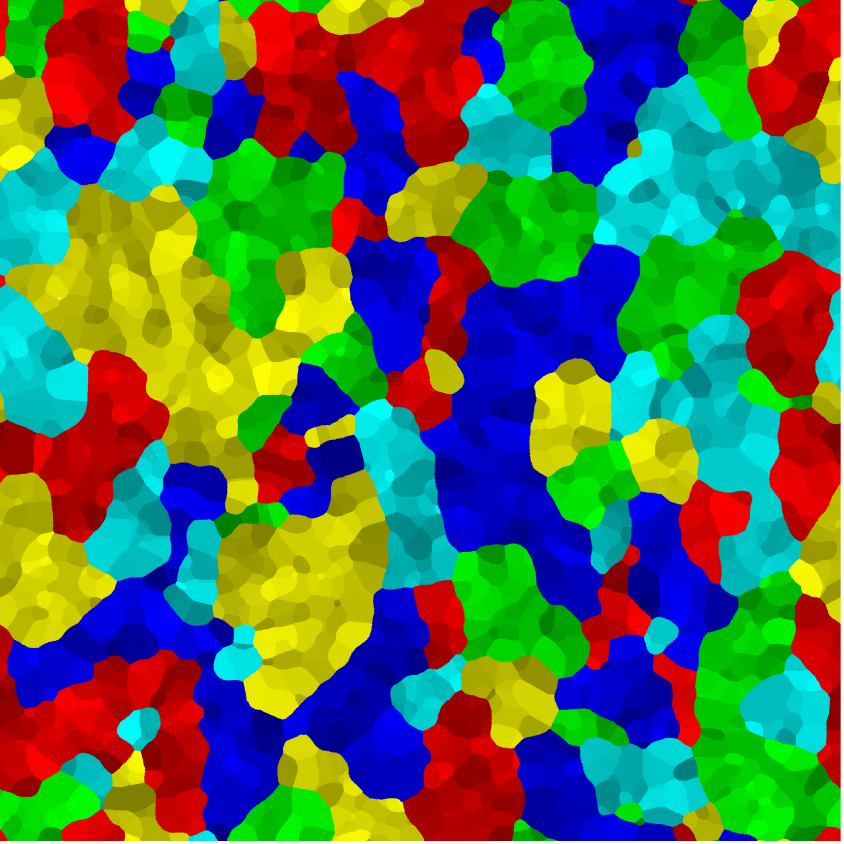}
            \caption{$t=500t_S$}   
            \label{subfig:subgrained_crystal_t500}
        \end{subfigure}
        \begin{subfigure}[b]{0.32\textwidth}  
            \centering 
            \includegraphics[width=\textwidth]{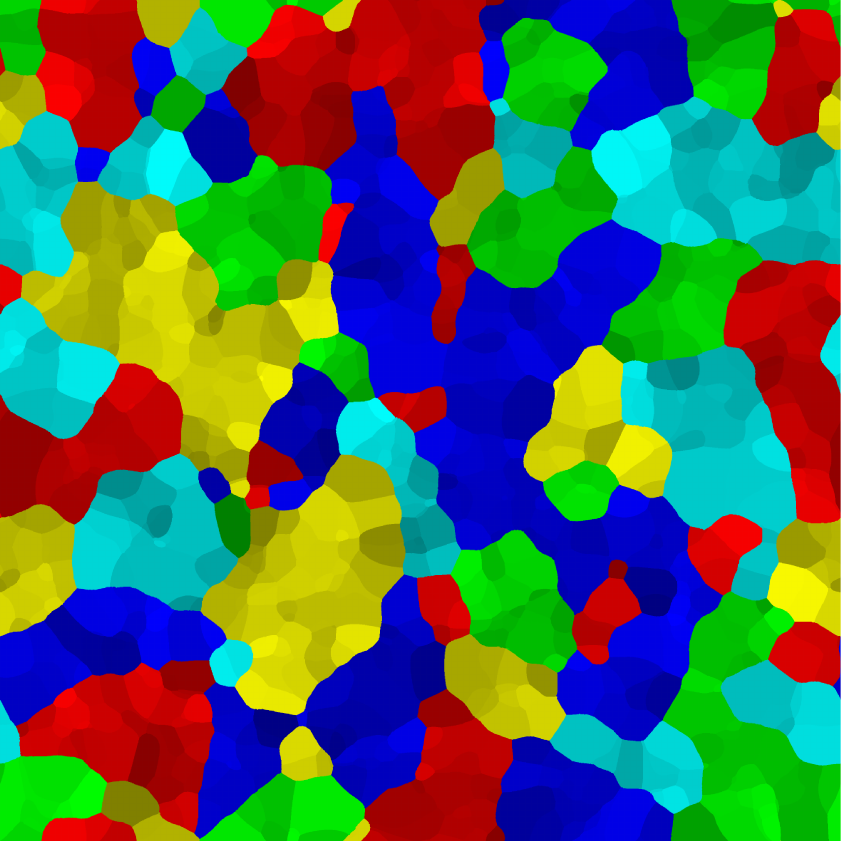}
            \caption{$t=2000t_S$}   
            \label{subfig:subgrained_crystal_t2000}
        \end{subfigure}
        \caption{Microstructure evolution of a polycrystal consisting of subgrains (same colored grains). The boundaries between groups A (red) and E(blue) are twin grain boundaries.}
        \label{fig:subgrained_polycrystal}
\end{figure}
%LAGB statistics 
\begin{figure}
        \centering
        \begin{subfigure}[b]{0.32\textwidth}
            \centering
        \includegraphics[width=\textwidth]{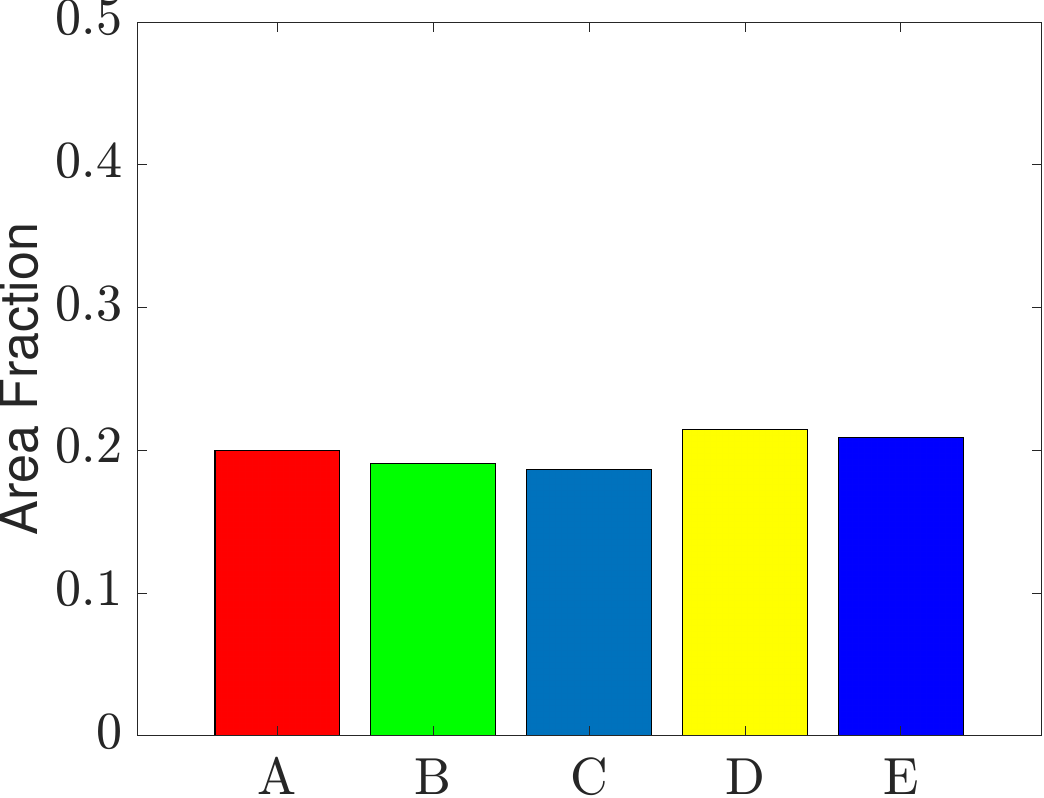}
            \caption{Initial condition}
            \label{subfig:texture_LAGB_t000}
        \end{subfigure}
        \begin{subfigure}[b]{0.32\textwidth}  
            \centering 
            \includegraphics[width=\textwidth]{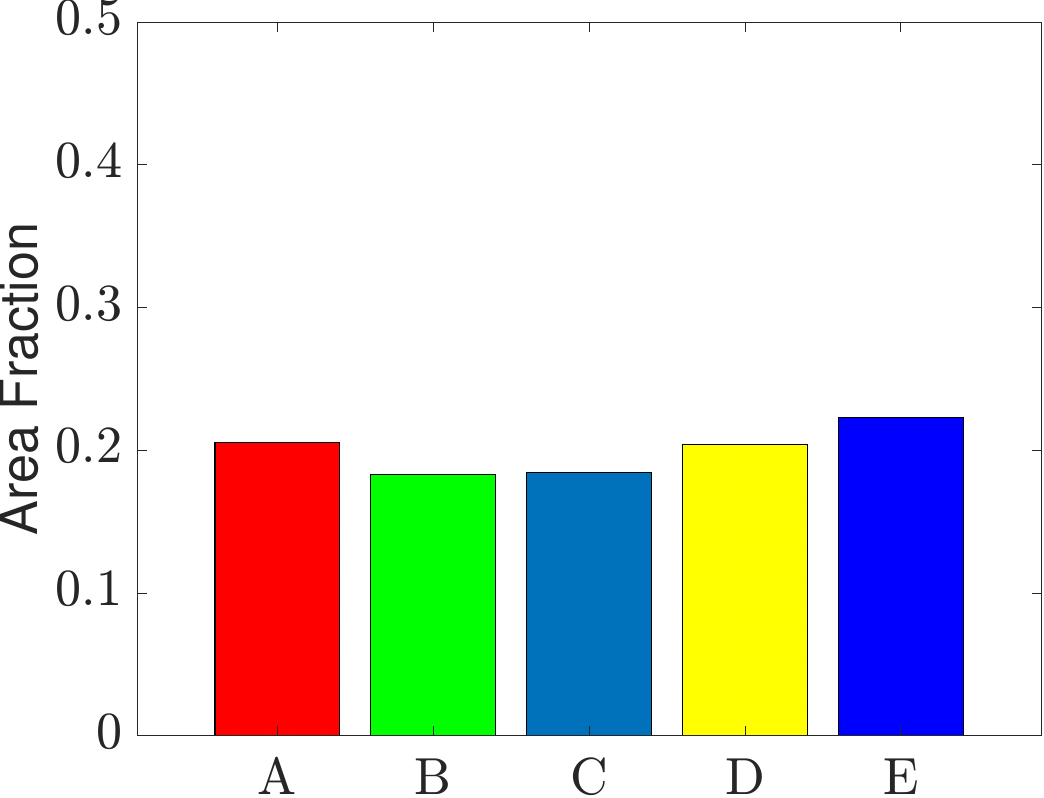}
            \caption{$t=500t_S$}   
            \label{subfig:texture_LAGB_t500}
        \end{subfigure}
        \begin{subfigure}[b]{0.32\textwidth}  
            \centering 
            \includegraphics[width=\textwidth]{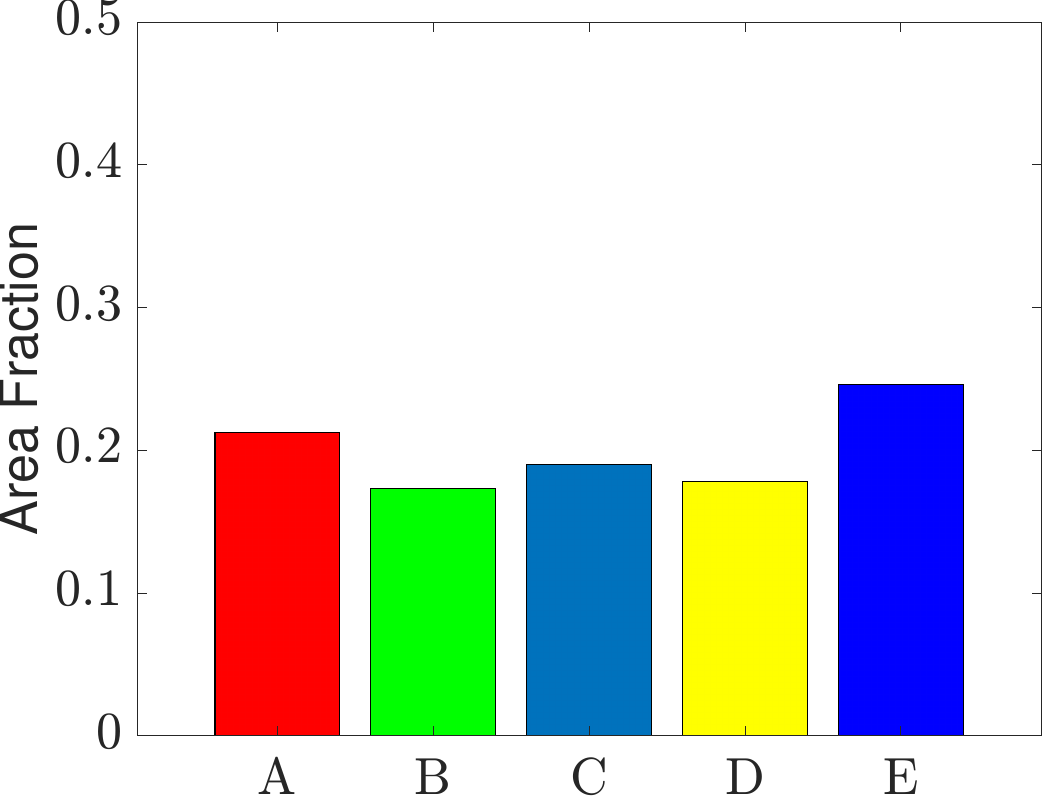}
            \caption{$t=2000t_S$}   
            \label{subfig:texture_LAGB_t2000}
        \end{subfigure}
        \vspace{0.5mm}
                \begin{subfigure}[b]{0.32\textwidth}
            \centering
        \includegraphics[width=\textwidth]{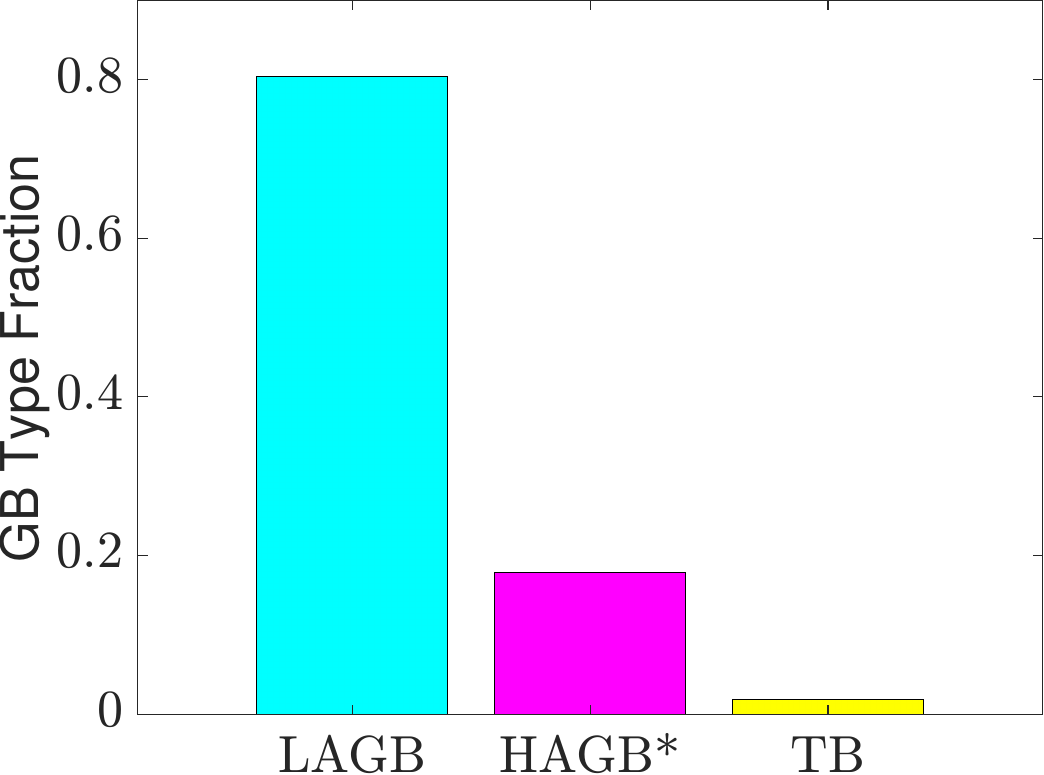}
            \caption{Initial condition}
            \label{subfig:gbtype_LAGB_ic}
        \end{subfigure}
        \begin{subfigure}[b]{0.32\textwidth}  
            \centering 
            \includegraphics[width=\textwidth]{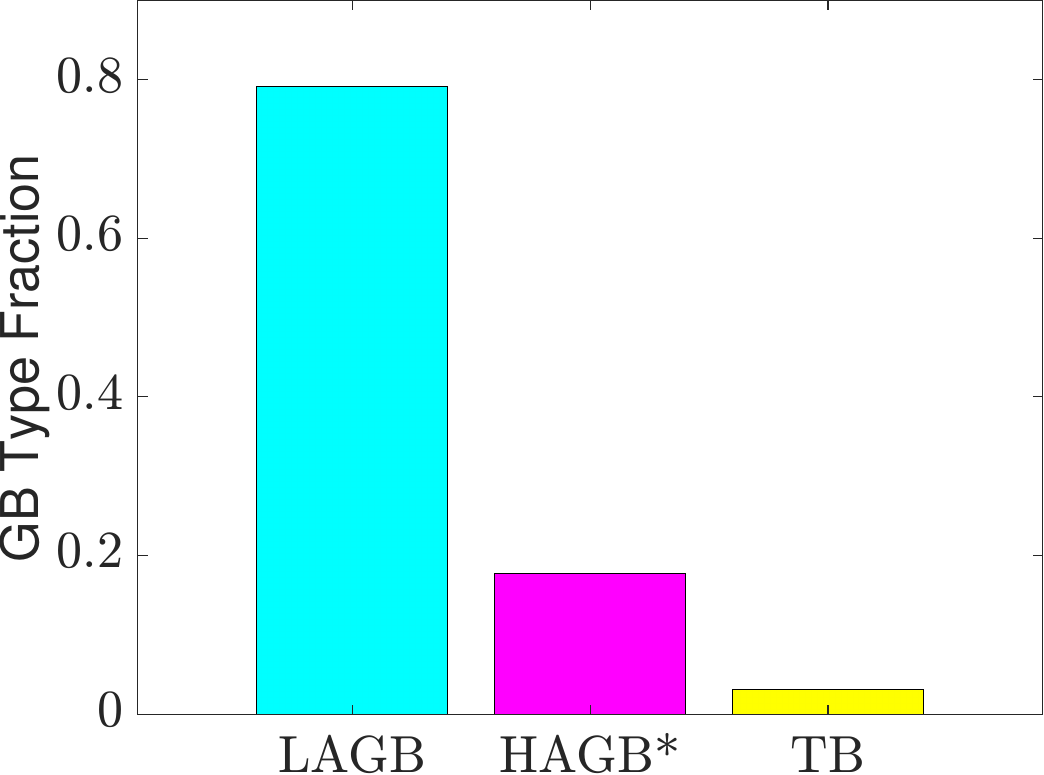}
            \caption{$t=500t_S$}   
            \label{subfig:gbtype_LAGB_t500}
        \end{subfigure}
        \begin{subfigure}[b]{0.32\textwidth}  
            \centering 
            \includegraphics[width=\textwidth]{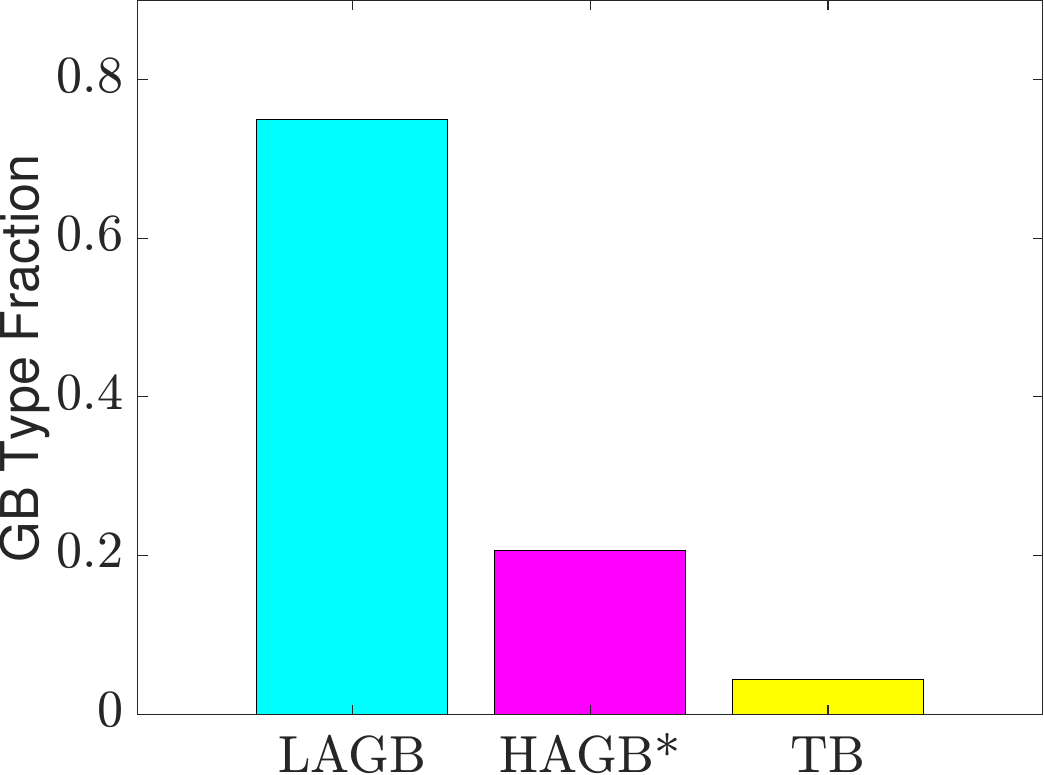}
            \caption{$t=2000t_S$}   
            \label{subfig:gbtype_LAGB_t2000}
        \end{subfigure}
        \caption{Texture (\protect\subref{subfig:texture_LAGB_t500}-\protect\subref{subfig:texture_LAGB_t2000}) and 
        types of grain boundary (\protect\subref{subfig:gbtype_LAGB_t500}-\protect\subref{subfig:gbtype_LAGB_t2000}) changes during grain growth of a LAGB dominant polycrystal. HAGB* refers high-angle grain boundaries excluding TBs. }
        \label{fig:gstatistics_LAGB}
\end{figure}

\fref{subfig:subgrained_crystal_ic} through \fref{subfig:subgrained_crystal_t2000} 
show the time evolution of a grain microstructure with subgrains and a high
fraction of LAGBs. Similar to Case 2a, certain subgrains located close to a GB
begin to grow at the expense of subgrains of the adjoining grain. To examine how
the statistics of the microstructure evolve, we classified the grains into three
types: LAGB, HAGB*, and TB, where HAGB* represents high-angle grain boundaries
that are not TBs. \fref{subfig:texture_LAGB_t000} to
\fref{subfig:texture_LAGB_t2000} show plots of the area fraction of each group (A to
E) and \fref{subfig:gbtype_LAGB_ic} to \fref{subfig:gbtype_LAGB_t2000} show the
fraction of GB types, at three different times. 
The plot in \fref{subfig:gbtype_LAGB_ic} shows that the majority of GBs
in the initial microstructure are LAGBs, and only a negligible fraction of them
are TBs. Since TBs have low energy, we expect area fractions of grain
types A and E --- which together form a TB --- to grow. However, we observe that there
is only a marginal increase in their area fractions in the
final microstructure (\fref{subfig:texture_LAGB_t2000}).
The increase in the fraction of TBs is also negligible. Based on the above
observation, we conclude that in a system with a high fraction of LAGBs, there
is no strong preference for TBs and texture does not form.
 
\begin{figure}
        \centering
        \begin{subfigure}[b]{0.32\textwidth}
            \centering
        \includegraphics[width=\textwidth]{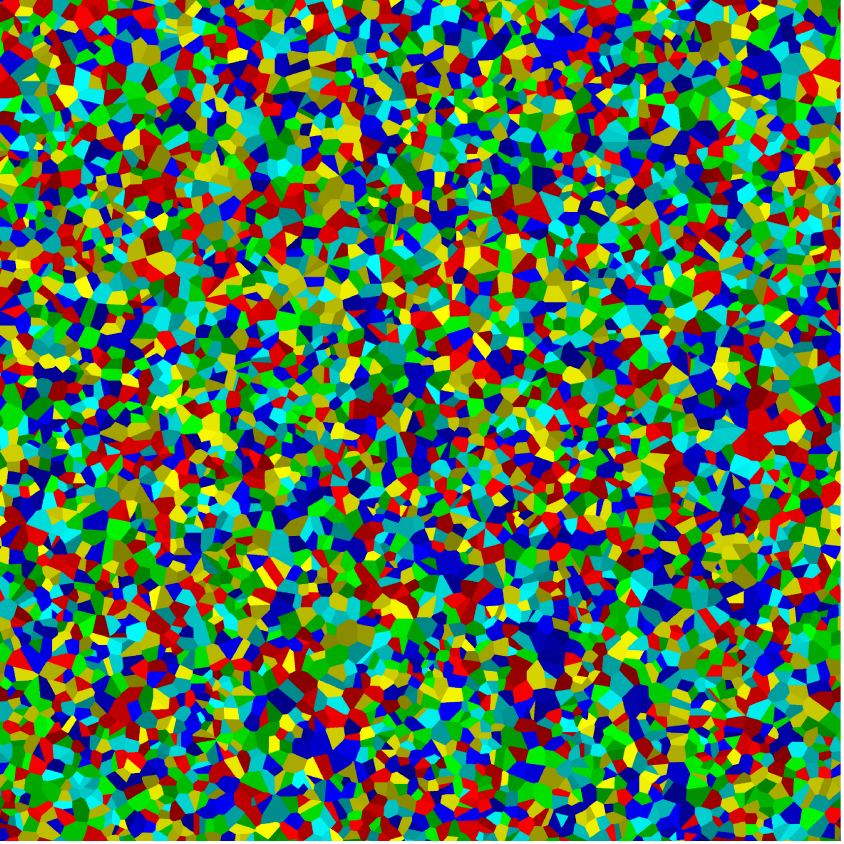}
            \caption{Initial condition}
            \label{subfig:grains_HAGB_ic}
        \end{subfigure}
        \begin{subfigure}[b]{0.32\textwidth}  
            \centering 
            \includegraphics[width=\textwidth]{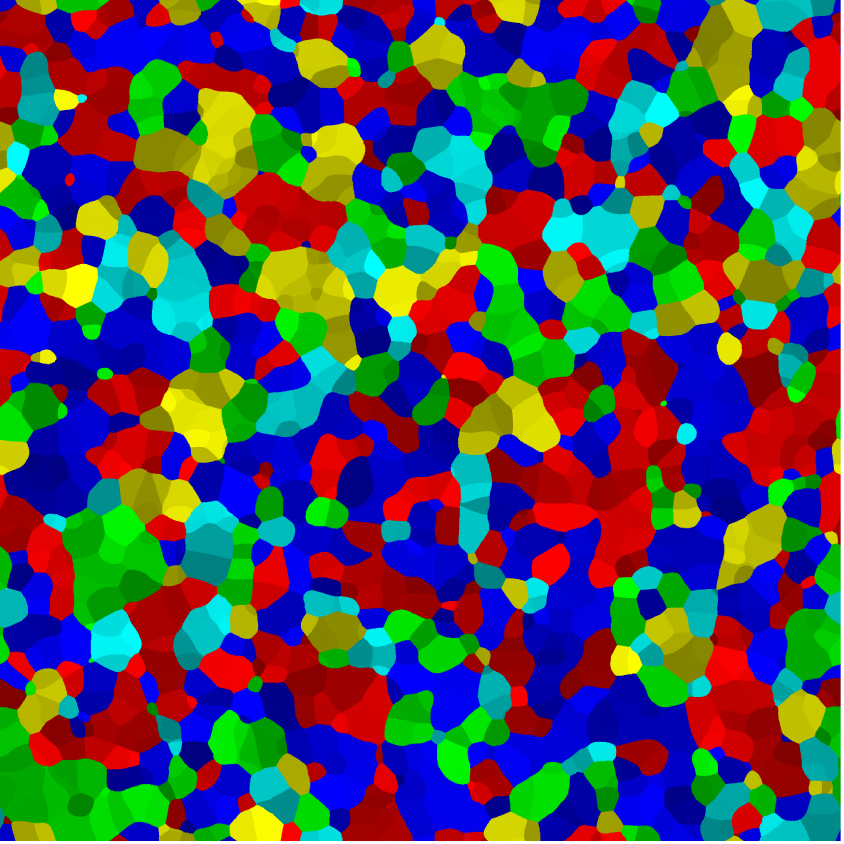}
            \caption{$t=500t_S$}   
            \label{subfig:grains_HAGB_t500}
        \end{subfigure}
        \begin{subfigure}[b]{0.32\textwidth}  
            \centering 
            \includegraphics[width=\textwidth]{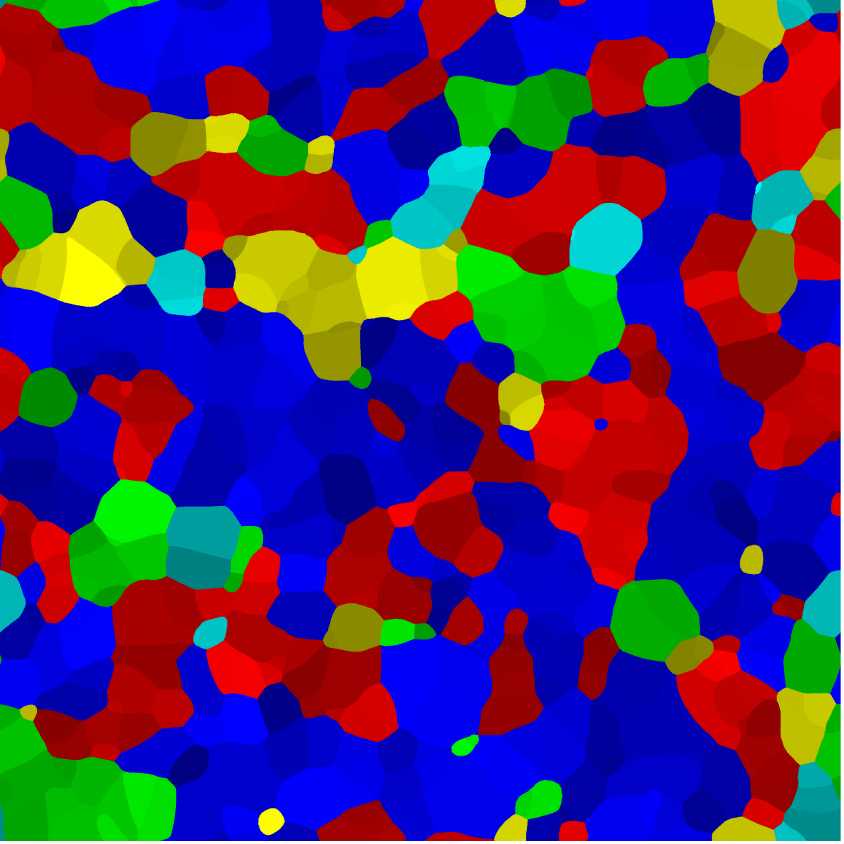}
            \caption{$t=2000t_S$}   
            \label{subfig:grains_HAGB_t2000}
        \end{subfigure}
        \caption{Microstructure evolution of a HAGB dominant polycrystal~ (\protect\subref{subfig:grains_HAGB_ic}). The boundaries between groups A (red) and E(blue) are twin grain boundaries. Grains from the same group coalesce first (\protect\subref{subfig:grains_HAGB_t500}) and type A and E grains become dominant  (\protect\subref{subfig:grains_HAGB_t2000}). }
        \label{fig:grains_HAGB}
\end{figure}

%HAGB statistics 
\begin{figure}
        \centering
        \begin{subfigure}[b]{0.32\textwidth}
            \centering
        \includegraphics[width=\textwidth]{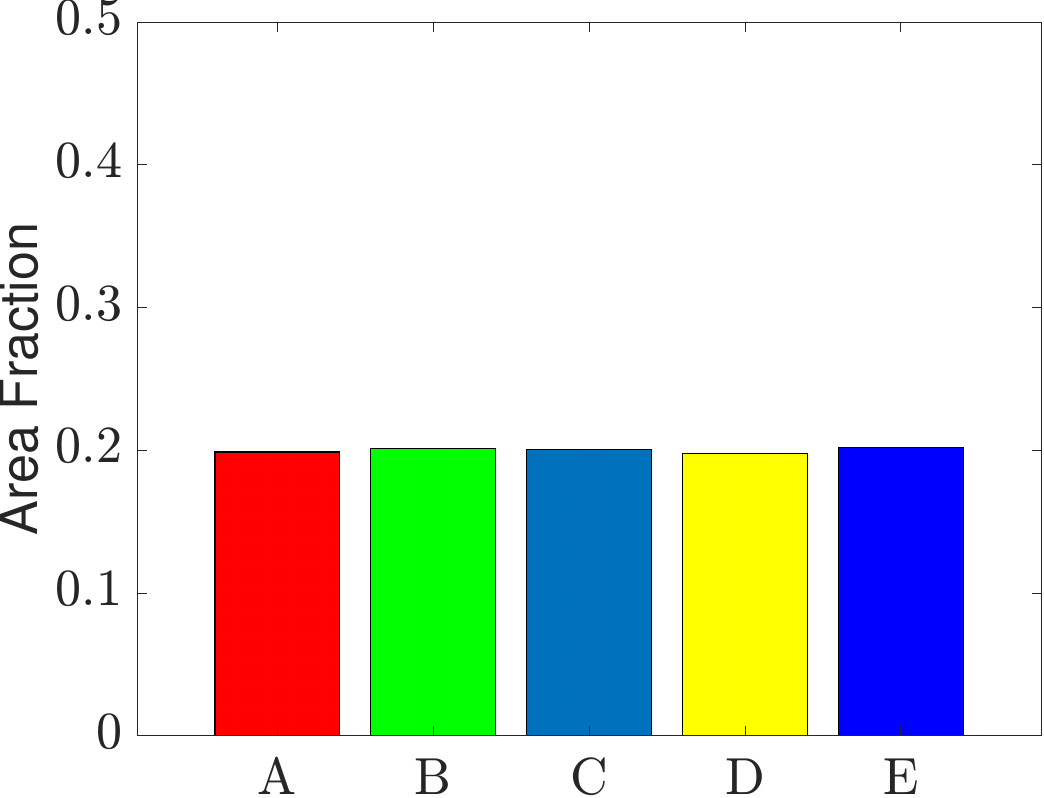}
            \caption{Initial condition}
            \label{subfig:texture_HAGB_t000}
        \end{subfigure}
        \begin{subfigure}[b]{0.32\textwidth}  
            \centering 
            \includegraphics[width=\textwidth]{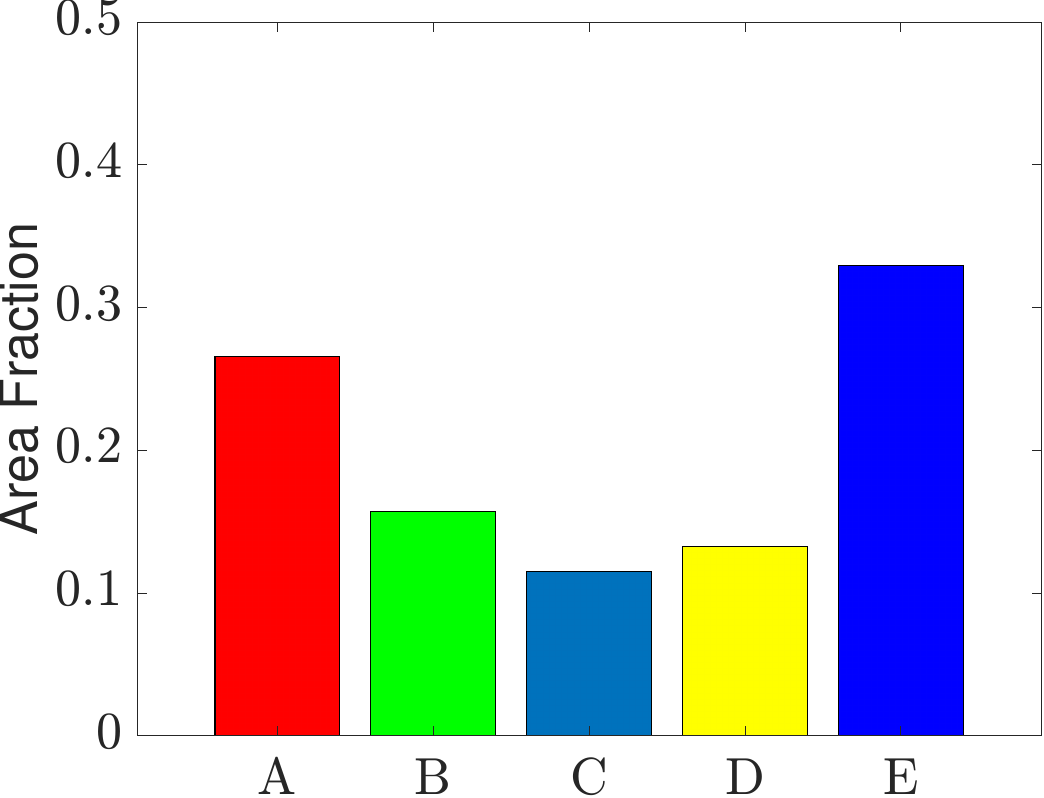}
            \caption{$t=500t_S$}   
            \label{subfig:texture_HAGB_t500}
        \end{subfigure}
        \begin{subfigure}[b]{0.32\textwidth}  
            \centering 
            \includegraphics[width=\textwidth]{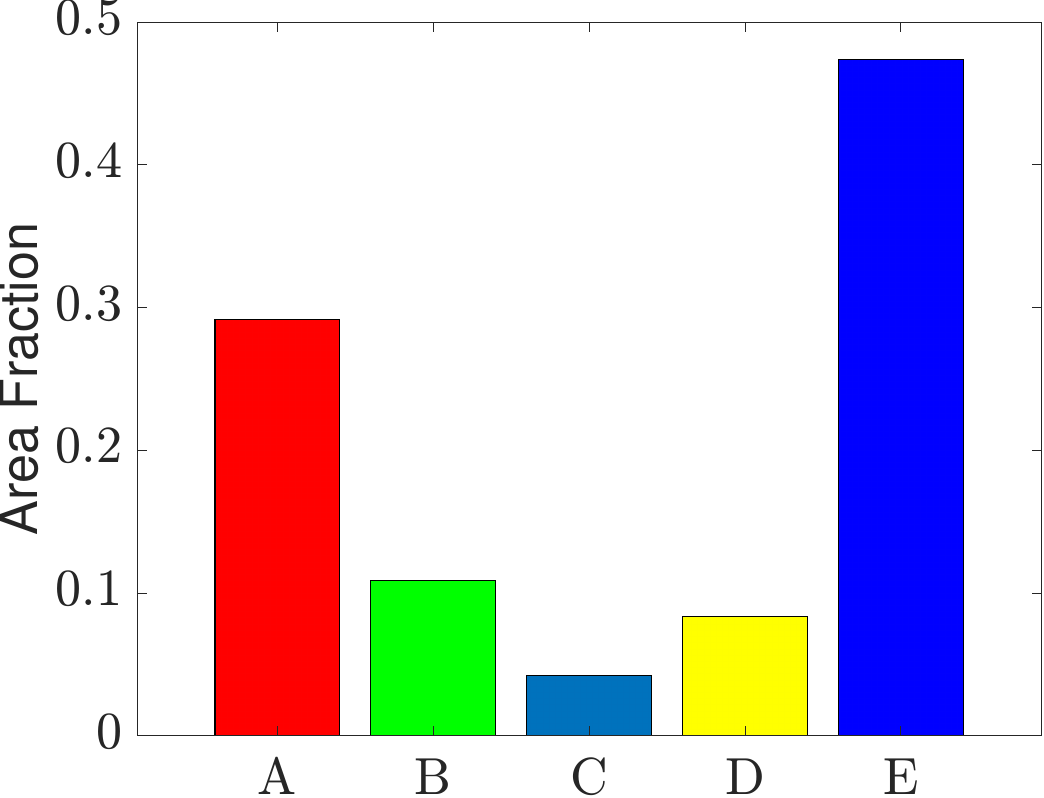}
            \caption{$t=2000t_S$}   
            \label{subfig:texture_HAGB_t2000}
        \end{subfigure}
        \vspace{0.5mm}
                \begin{subfigure}[b]{0.32\textwidth}
            \centering
        \includegraphics[width=\textwidth]{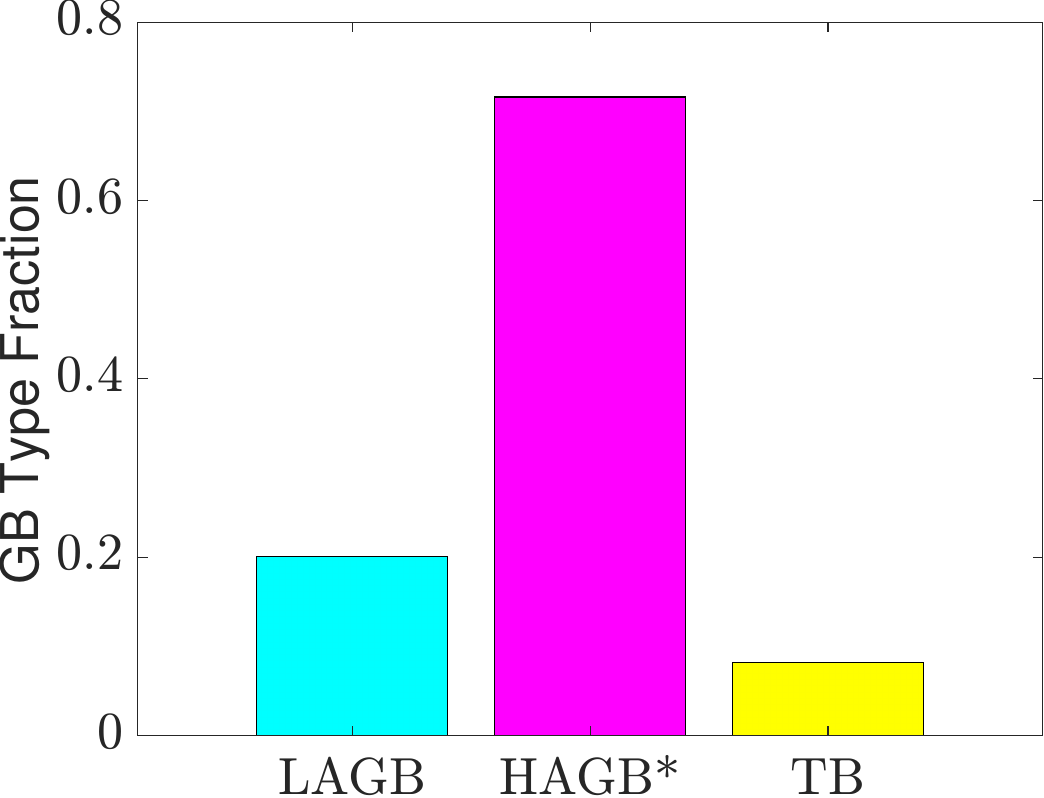}
            \caption{Initial condition}
            \label{subfig:gbtype_HAGB_ic}
        \end{subfigure}
        \begin{subfigure}[b]{0.32\textwidth}  
            \centering 
            \includegraphics[width=\textwidth]{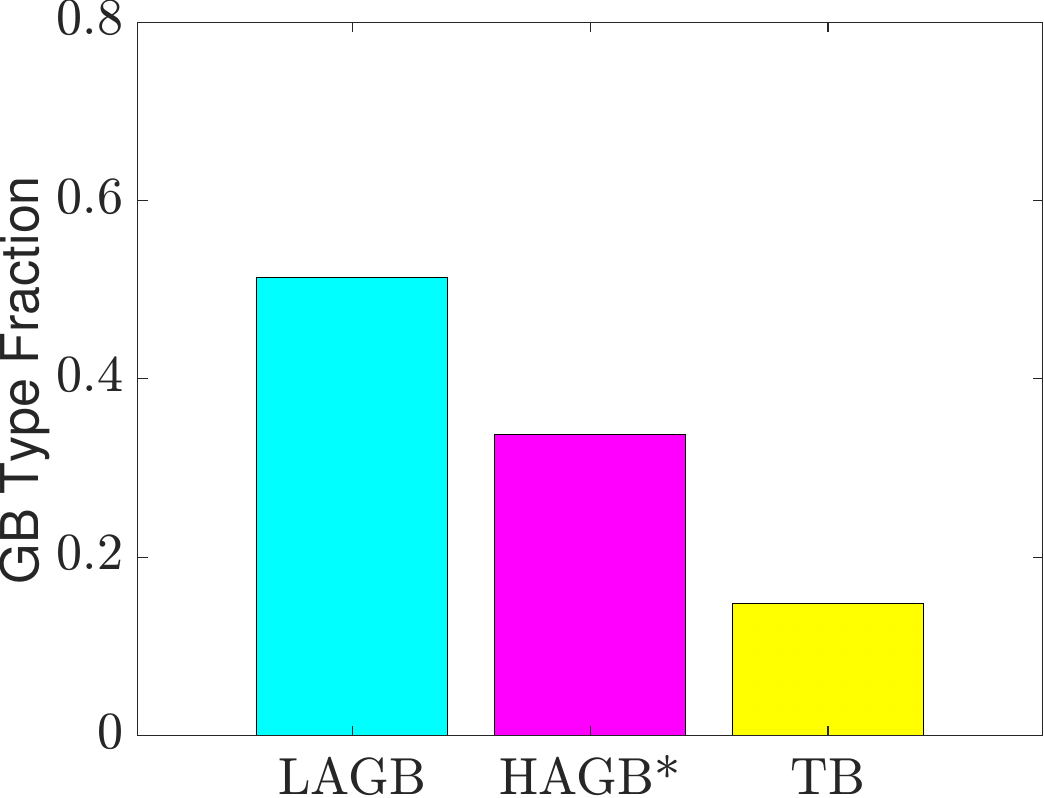}
            \caption{$t=500t_S$}   
            \label{subfig:gbtype_HAGB_t500}
        \end{subfigure}
        \begin{subfigure}[b]{0.32\textwidth}  
            \centering 
            \includegraphics[width=\textwidth]{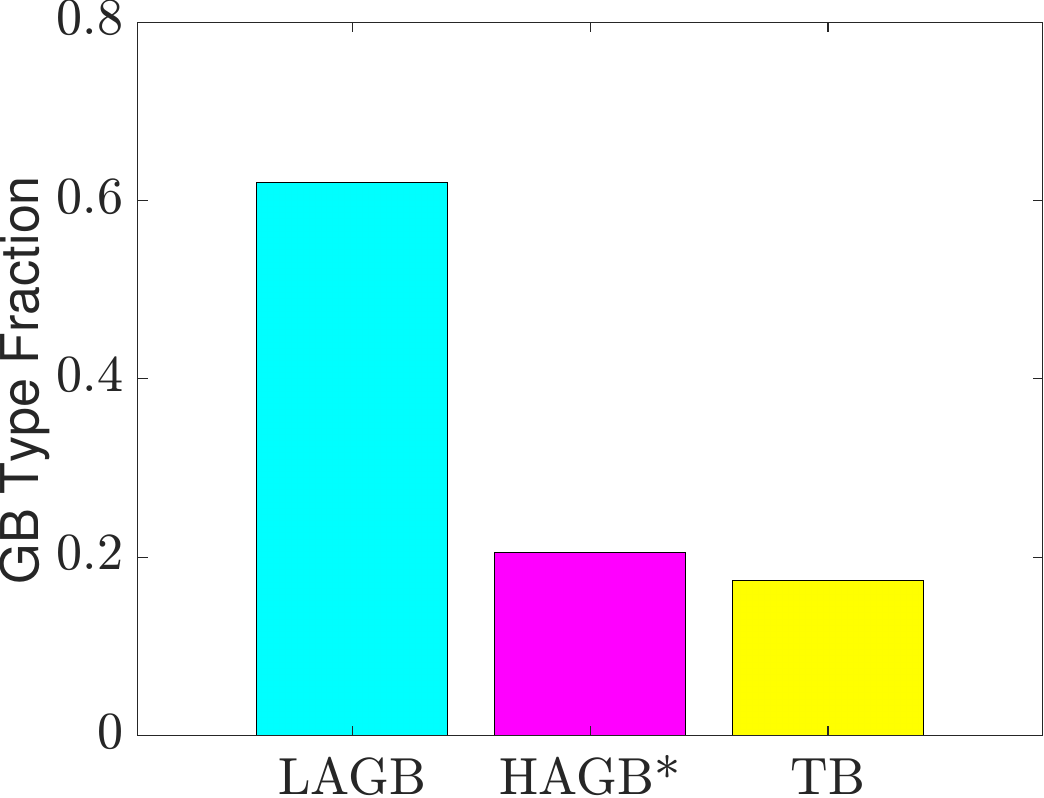}
            \caption{$t=2000t_S$}   
            \label{subfig:gbtype_HAGB_t2000}
        \end{subfigure}
        \caption{Texture (\protect\subref{subfig:texture_HAGB_t500}-\protect\subref{subfig:texture_HAGB_t2000}) and 
        types of grain boundary (\protect\subref{subfig:gbtype_HAGB_t500}-\protect\subref{subfig:gbtype_HAGB_t2000}) changes during grain growth of  a HAGB dominant polycrystal. HAGB* refers to high-angle grain boundaries excluding TBs. }
        \label{fig:gstatistics_HAGB}
\end{figure}

Next, we examine the microstructure evolution in Case 2c wherein HAGBs are
dominant and the system has no subgrains. Expectedly, the initial microstructure shown in
\fref{subfig:grains_HAGB_ic} has no subgrains. 
The initial fraction of TB is also higher than Case 2b (\fref{subfig:gbtype_HAGB_ic}).
However, the initial texture, seen in \fref{subfig:texture_HAGB_t000}, is not as strong as previous. 
In this case, as grain growth continues, grains from the same groups coalesce (\fref{subfig:grains_HAGB_t500}), 
because the system penalizes HAGB*, which have higher energies. 
During this process, the fractions of both LAGB and TBs steeply increase at the expense of HAGB*. Consequently, the final microstructure (\fref{subfig:grains_HAGB_t2000}) shows 
a considerable growth of type A and E grains, indicating texture development in the microstructure.
In \fref{subfig:gbtype_HAGB_t2000}, we also observed that the growth of TBs is substantially facilitated using an initial microstructure with a high fraction of HAGB. 

Lastly, we discuss the results of the present simulations in comparison to
recent experimental observations of TB development in nanocrystalline
materials~\citep{David:2021}. \citet{David:2021} investigated the effect of
grain size on texture formation during
annealing  using electro-deposited nickel samples
with different average grain sizes, ranging from 20 nm to  200 nm. It is
observed that the fraction of TBs in the final microstructure
increases as the average grain size in the initial microstructure decreases. 
This trend is attributed to the relationship between the probability of
accidental twin formation and the velocity of the migrating grain boundary,
which in turn is inversely proportional to grain size. 
In our simulation, the initial microstructure of Case 2c can also be viewed as a
substantially smaller average grain size than that of Case 2b (if one ignores
its sub-microstructure). Since smaller grains coarsen faster, the grains in Case
2c have a higher probability of forming new grain boundaries during the same
period of time, which also increases the probability of forming twin grain
boundaries. In this regard, our simulation results of Case 2c are consistent with experimental
observations. Yet, it is also important to note that the size difference is not
the only factor in our simulations. The main difference between cases 2b and 2c
is the fraction of grain boundary types in the initial condition. Unfortunately,
however, since the initial fractions of GB types are not reported in
Ref.~\citep{David:2021}, a conclusive discussion cannot be made at this time.

\section{Conclusion}
\label{sec:conclusion}

A fundamental open problem in materials science is to establish the relationship between process parameters and the evolution of grain microstructure.
Given that the process-structure relationship is inherently statistical, it is necessary to use lightweight models that can efficiently capture the microstructure evolution of a polycrystal ensemble.
Recent years have seen considerable advances in threshold-dynamics techniques, which have revolutionized the way to simulate full-field grain microstructure evolution during growth. 
The method serves as a highly efficient and robust algorithm for statistical studies on grain microstructure. 

In this paper, we utilized the TD method to investigate the statistical behavior of grain microstructure under anisotropic GB characters, with a focus on abnormal grain growth and texture development. To ensure the numerical stability of the algorithm and reliable result dynamic evolution of the grain network, we imposed a restriction on the degree of GB anisotropy. We considered GBs with energies and mobilities that are compatible with the fundamental restrictions of the threshold-dynamics method. 

Our first numerical experiment involves a system with the simplest GB anisotropy, which facilitates the analysis of simulations of abnormal grain growth. We found that GB anisotropy introduces a statistical preference for certain grain orientations, leading to changes to the grain size distribution compared to an isotropic system.
In our second numerical experiment, we incorporated crystallographic grain boundary energy and examined the evolution of microstructure features at different initial configurations. We observed that the development of texture and the growth of twin grain boundaries were more pronounced when the initial microstructure had a higher fraction of high-angle grain boundaries. These findings suggest effective grain boundary engineering strategies for improving material properties. 

In our future work, we aim to enhance the TD method by integrating grain rotation~\citep{Selim_Grotation} and grain boundary plasticity~\citep{Admal:2018}, allowing for the simultaneous evolution of microstructure and deformation. Especially, this is crucial for investigating phenomena such as dynamic recrystallization, superplasticity, and severe plastic deformation~\citep{Srolovitz:2017,Srolovitz:2020,Runnels:2020}, which require a more comprehensive understanding of the underlying mechanisms. 
We anticipate that incorporating these features will enhance the accuracy and applicability of the model, thereby advancing our ability to predict and optimize material properties.

\section*{Declaration of competing interest}

The authors declare that they have no known competing financial interests or personal relationships that could have appeared to influence the work reported in this paper.

\bibliographystyle{model1-num-names}
\bibliography{references}

\begin{thebibliography}{68}
\expandafter\ifx\csname natexlab\endcsname\relax\def\natexlab#1{#1}\fi
\providecommand{\url}[1]{\texttt{#1}}
\providecommand{\href}[2]{#2}
\providecommand{\path}[1]{#1}
\providecommand{\DOIprefix}{doi:}
\providecommand{\ArXivprefix}{arXiv:}
\providecommand{\URLprefix}{URL: }
\providecommand{\Pubmedprefix}{pmid:}
\providecommand{\doi}[1]{\href{http://dx.doi.org/#1}{\path{#1}}}
\providecommand{\Pubmed}[1]{\href{pmid:#1}{\path{#1}}}
\providecommand{\bibinfo}[2]{#2}
\ifx\xfnm\relax \def\xfnm[#1]{\unskip,\space#1}\fi
%Type = Article
\bibitem[{Watanabe(2011)}]{Watanabe:2011}
\bibinfo{author}{T.~Watanabe},
\newblock \bibinfo{title}{Grain boundary engineering: historical perspective
  and future prospects},
\newblock \bibinfo{journal}{Journal of Material Science} \bibinfo{volume}{46}
  (\bibinfo{year}{2011}) \bibinfo{pages}{4095--4115}.
%Type = Article
\bibitem[{Kumar et~al.(2003)Kumar, Swygenhoven, and Suresh}]{NCmaterial01}
\bibinfo{author}{K.~S. Kumar}, \bibinfo{author}{H.~V. Swygenhoven},
  \bibinfo{author}{S.~Suresh},
\newblock \bibinfo{title}{Mechanical behavior of nanocrystalline metals and
  alloys},
\newblock \bibinfo{journal}{Acta Materialia} \bibinfo{volume}{51}
  (\bibinfo{year}{2003}) \bibinfo{pages}{5743--5774}.
%Type = Article
\bibitem[{II and Boyce(2010)}]{NCmaterial02}
\bibinfo{author}{H.~A.~P. II}, \bibinfo{author}{B.~L. Boyce},
\newblock \bibinfo{title}{A review of fatigue behavior in nanocrystalline
  metals},
\newblock \bibinfo{journal}{Experimental Mechanics} \bibinfo{volume}{50}
  (\bibinfo{year}{2010}) \bibinfo{pages}{5--23}.
%Type = Article
\bibitem[{Zhu and Langdon(2004)}]{NCmaterial03}
\bibinfo{author}{Y.~T. Zhu}, \bibinfo{author}{T.~G. Langdon},
\newblock \bibinfo{title}{The fundamentals of nanostructured materials
  processed by severe plastic deformation},
\newblock \bibinfo{journal}{The Journal of The Minerals, Metals \& Materials
  Society} \bibinfo{volume}{56} (\bibinfo{year}{2004}) \bibinfo{pages}{58--63}.
%Type = Article
\bibitem[{Bober et~al.(2015)Bober, Kumar, and Rupert}]{Rupert:2015}
\bibinfo{author}{D.~B. Bober}, \bibinfo{author}{M.~Kumar},
  \bibinfo{author}{T.~J. Rupert},
\newblock \bibinfo{title}{Nanocrystalline grain boundary engineering:
  Increasing ${\Sigma}$3 boundary fraction in pure {Ni} with thermomechanical
  treatments},
\newblock \bibinfo{journal}{Acta Materialia} \bibinfo{volume}{86}
  (\bibinfo{year}{2015}) \bibinfo{pages}{43--54}.
%Type = Article
\bibitem[{Palumbo et~al.(1998)Palumbo, Lehockey, and Lin}]{NCmaterial04}
\bibinfo{author}{G.~Palumbo}, \bibinfo{author}{E.~M. Lehockey},
  \bibinfo{author}{P.~Lin},
\newblock \bibinfo{title}{Applications for grain boundary engineered
  materials},
\newblock \bibinfo{journal}{The Journal of The Minerals, Metals \& Materials
  Society} \bibinfo{volume}{50} (\bibinfo{year}{1998}) \bibinfo{pages}{40--43}.
%Type = Article
\bibitem[{Frolov et~al.(2018)Frolov, Setyawan, Kurtz, Marian, Oganov, Rudd, and
  Zhu}]{Frolov:2018}
\bibinfo{author}{T.~Frolov}, \bibinfo{author}{W.~Setyawan},
  \bibinfo{author}{R.~J. Kurtz}, \bibinfo{author}{J.~Marian},
  \bibinfo{author}{A.~R. Oganov}, \bibinfo{author}{R.~E. Rudd},
  \bibinfo{author}{Q.~Zhu},
\newblock \bibinfo{title}{Grain boundary phases in bcc metals},
\newblock \bibinfo{journal}{Nanoscale} \bibinfo{volume}{10}
  (\bibinfo{year}{2018}) \bibinfo{pages}{8253--8268}.
%Type = Article
\bibitem[{Janssens et~al.(2006)Janssens, Olmsted, Holm, Foiles, Plimton, and
  Derlet}]{Foiles:2006}
\bibinfo{author}{K.~G.~F. Janssens}, \bibinfo{author}{D.~Olmsted},
  \bibinfo{author}{E.~A. Holm}, \bibinfo{author}{S.~M. Foiles},
  \bibinfo{author}{S.~J. Plimton}, \bibinfo{author}{P.~M. Derlet},
\newblock \bibinfo{title}{Computing the mobility of grain boundaries},
\newblock \bibinfo{journal}{Nature Materials} \bibinfo{volume}{5}
  (\bibinfo{year}{2006}) \bibinfo{pages}{124--127}.
%Type = Article
\bibitem[{Molodov and Molodov(2018)}]{MOLODOV2018336}
\bibinfo{author}{K.~D. Molodov}, \bibinfo{author}{D.~A. Molodov},
\newblock \bibinfo{title}{Grain boundary mediated plasticity: On the evaluation
  of grain boundary migration - shear coupling},
\newblock \bibinfo{journal}{Acta Materialia} \bibinfo{volume}{153}
  (\bibinfo{year}{2018}) \bibinfo{pages}{336--353}.
%Type = Article
\bibitem[{Thomas et~al.(2017)Thomas, Chen, Han, Purohit, and
  Srolovitz}]{Srolovitz:2017}
\bibinfo{author}{S.~Thomas}, \bibinfo{author}{K.~Chen},
  \bibinfo{author}{J.~Han}, \bibinfo{author}{P.~K. Purohit},
  \bibinfo{author}{D.~J. Srolovitz},
\newblock \bibinfo{title}{Reconciling grain growth and shear-coupled grain
  boundary migration},
\newblock \bibinfo{journal}{Nature communications} \bibinfo{volume}{8}
  (\bibinfo{year}{2017}) \bibinfo{pages}{1--12}.
%Type = Article
\bibitem[{Kim et~al.(2021)Kim, Jacobs, Osher, and Admal}]{JKIM:2021}
\bibinfo{author}{J.~Kim}, \bibinfo{author}{M.~Jacobs},
  \bibinfo{author}{S.~Osher}, \bibinfo{author}{N.~C. Admal},
\newblock \bibinfo{title}{A crystal symmetry-invariant
  {K}obayashi--{W}arren--{C}arter grain boundary model and its implementation
  using a thresholding algorithm},
\newblock \bibinfo{journal}{Computational Materials Science}
  \bibinfo{volume}{199} (\bibinfo{year}{2021}) \bibinfo{pages}{110575}.
%Type = Article
\bibitem[{Belytschko et~al.(2009)Belytschko, Gracie, and
  Ventura}]{Belytschko:2009}
\bibinfo{author}{T.~Belytschko}, \bibinfo{author}{R.~Gracie},
  \bibinfo{author}{G.~Ventura},
\newblock \bibinfo{title}{A review of extended/generalized finite element
  methods for material modeling},
\newblock \bibinfo{journal}{Modelling and Simulation in Materials Science and
  Engineering} \bibinfo{volume}{17} (\bibinfo{year}{2009})
  \bibinfo{pages}{043001}.
%Type = Article
\bibitem[{Zhao et~al.(2022)Zhao, Wahab, Ling, and Liu}]{ZHAO2022275}
\bibinfo{author}{Q.~Zhao}, \bibinfo{author}{M.~A. Wahab},
  \bibinfo{author}{Y.~Ling}, \bibinfo{author}{Z.~Liu},
\newblock \bibinfo{title}{Fatigue crack propagation across grain boundary of
  {Al-Cu-Mg} bicrystal based on crystal plasticity {XFEM} and cohesive zone
  model},
\newblock \bibinfo{journal}{Journal of Materials Science \& Technology}
  \bibinfo{volume}{126} (\bibinfo{year}{2022}) \bibinfo{pages}{275--287}.
%Type = Article
\bibitem[{Liu et~al.(2014)Liu, He, Yao, Li, and Tang}]{LIU2014310}
\bibinfo{author}{W.~Liu}, \bibinfo{author}{Z.~He}, \bibinfo{author}{W.~Yao},
  \bibinfo{author}{M.~Li}, \bibinfo{author}{J.~Tang},
\newblock \bibinfo{title}{{XFEM} simulation of the effects of microstructure on
  the intergranular fracture in high strength aluminum alloy},
\newblock \bibinfo{journal}{Computational Materials Science}
  \bibinfo{volume}{84} (\bibinfo{year}{2014}) \bibinfo{pages}{310--317}.
%Type = Article
\bibitem[{Mullins(1956)}]{Mullins}
\bibinfo{author}{W.~W. Mullins},
\newblock \bibinfo{title}{Two-dimensional motion of idealized grain
  boundaries},
\newblock \bibinfo{journal}{Journal of Applied Physics} \bibinfo{volume}{27}
  (\bibinfo{year}{1956}) \bibinfo{pages}{900--904}.
%Type = Article
\bibitem[{Barmak et~al.(2013)Barmak, Eggeling, Kinderlehrer, Sharp, Ta'asan,
  Rollett, and Coffey}]{Barmak:2013}
\bibinfo{author}{K.~Barmak}, \bibinfo{author}{E.~Eggeling},
  \bibinfo{author}{D.~Kinderlehrer}, \bibinfo{author}{R.~Sharp},
  \bibinfo{author}{S.~Ta'asan}, \bibinfo{author}{A.~D. Rollett},
  \bibinfo{author}{K.~R. Coffey},
\newblock \bibinfo{title}{Grain growth and the puzzle of its stagnation in thin
  films: The curious tale of a tail and an ear},
\newblock \bibinfo{journal}{Progress in Materials Science}
  (\bibinfo{year}{2013}) \bibinfo{pages}{987--1055}.
%Type = Article
\bibitem[{Lazar et~al.(2020)Lazar, Mason, MacPherson, and
  Srolovitz}]{Lazar:2020}
\bibinfo{author}{E.~A. Lazar}, \bibinfo{author}{J.~K. Mason},
  \bibinfo{author}{R.~D. MacPherson}, \bibinfo{author}{D.~J. Srolovitz},
\newblock \bibinfo{title}{Distribution of topological types in grain-growth
  microstructures},
\newblock \bibinfo{journal}{Physical Review Letters} \bibinfo{volume}{125}
  (\bibinfo{year}{2020}) \bibinfo{pages}{015501}.
%Type = Article
\bibitem[{Mohles(2020)}]{Mohles}
\bibinfo{author}{V.~Mohles},
\newblock \bibinfo{title}{3-{D} front tracking model for interfaces with
  anisotropic energy},
\newblock \bibinfo{journal}{Computational Materials Science}
  \bibinfo{volume}{176} (\bibinfo{year}{2020}) \bibinfo{pages}{109534}.
%Type = Article
\bibitem[{Kim et~al.(2014)Kim, Kim, Dong, Steinbach, and Lee}]{BLEE:2014}
\bibinfo{author}{H.-K. Kim}, \bibinfo{author}{S.~G. Kim},
  \bibinfo{author}{W.~Dong}, \bibinfo{author}{I.~Steinbach},
  \bibinfo{author}{B.-J. Lee},
\newblock \bibinfo{title}{Phase-field modeling for 3d grain growth based on a
  grain boundary energy database},
\newblock \bibinfo{journal}{Modelling and Simulation in Materials Science and
  Engineering} \bibinfo{volume}{22} (\bibinfo{year}{2014})
  \bibinfo{pages}{034004}.
%Type = Article
\bibitem[{Salama et~al.(2020)Salama, Kundin, Shchyglo, Mohles, Marquardt, and
  Steinbach}]{SALAMA2020641}
\bibinfo{author}{H.~Salama}, \bibinfo{author}{J.~Kundin},
  \bibinfo{author}{O.~Shchyglo}, \bibinfo{author}{V.~Mohles},
  \bibinfo{author}{K.~Marquardt}, \bibinfo{author}{I.~Steinbach},
\newblock \bibinfo{title}{Role of inclination dependence of grain boundary
  energy on the microstructure evolution during grain growth},
\newblock \bibinfo{journal}{Acta Materialia} \bibinfo{volume}{188}
  (\bibinfo{year}{2020}) \bibinfo{pages}{641--651}.
%Type = Article
\bibitem[{Chen et~al.(2020)Chen, Han, Pan, and Srolovitz}]{Srolovitz:2020_2}
\bibinfo{author}{K.~Chen}, \bibinfo{author}{J.~Han}, \bibinfo{author}{X.~Pan},
  \bibinfo{author}{D.~J. Srolovitz},
\newblock \bibinfo{title}{The grain boundary mobility tensor},
\newblock \bibinfo{journal}{Proceedings of the National Academy of Sciences of
  the United States of America} \bibinfo{volume}{117} (\bibinfo{year}{2020})
  \bibinfo{pages}{4533--4538}.
%Type = Article
\bibitem[{Runnels et~al.(2016{\natexlab{a}})Runnels, Beyerlein, Conti, and
  Ortiz}]{Runnels:2016_1}
\bibinfo{author}{B.~Runnels}, \bibinfo{author}{I.~J. Beyerlein},
  \bibinfo{author}{S.~Conti}, \bibinfo{author}{M.~Ortiz},
\newblock \bibinfo{title}{An analytical model of interfacial energy based on a
  lattice-matching interatomic energy},
\newblock \bibinfo{journal}{Journal of Mechanics and Physics of Solids}
  \bibinfo{volume}{89} (\bibinfo{year}{2016}{\natexlab{a}})
  \bibinfo{pages}{174--193}.
%Type = Article
\bibitem[{Runnels et~al.(2016{\natexlab{b}})Runnels, Beyerlein, Conti, and
  Ortiz}]{Runnels:2016_2}
\bibinfo{author}{B.~Runnels}, \bibinfo{author}{I.~J. Beyerlein},
  \bibinfo{author}{S.~Conti}, \bibinfo{author}{M.~Ortiz},
\newblock \bibinfo{title}{A relaxation method for the energy and morphology of
  grain boundaries and interfaces},
\newblock \bibinfo{journal}{Journal of Mechanics and Physics of Solids}
  \bibinfo{volume}{94} (\bibinfo{year}{2016}{\natexlab{b}})
  \bibinfo{pages}{388--408}.
%Type = Article
\bibitem[{Olmsted et~al.(2009)Olmsted, Foiles, and Holm}]{Olmsted:2009}
\bibinfo{author}{D.~L. Olmsted}, \bibinfo{author}{S.~M. Foiles},
  \bibinfo{author}{E.~A. Holm},
\newblock \bibinfo{title}{Survey of computed grain boundary properties in
  face-centered cubic metals: I. grain boundary energy},
\newblock \bibinfo{journal}{Acta Materialla} \bibinfo{volume}{57}
  (\bibinfo{year}{2009}) \bibinfo{pages}{3694--3703}.
%Type = Article
\bibitem[{Hallberg and Bulatov(2019)}]{Bulatov:2019}
\bibinfo{author}{H.~Hallberg}, \bibinfo{author}{V.~V. Bulatov},
\newblock \bibinfo{title}{Modeling of grain growth under fully anisotropic
  grain boundary energy},
\newblock \bibinfo{journal}{Modelling and Simulation in Materials Science and
  Engineering} \bibinfo{volume}{27} (\bibinfo{year}{2019})
  \bibinfo{pages}{045002}.
%Type = Article
\bibitem[{Kalidindi and De~Graef(2015)}]{kalidindi2015materials}
\bibinfo{author}{S.~R. Kalidindi}, \bibinfo{author}{M.~De~Graef},
\newblock \bibinfo{title}{{Materials Data Science: Current Status and Future
  Outlook}},
\newblock \bibinfo{journal}{Annual Review of Materials Research}
  \bibinfo{volume}{45} (\bibinfo{year}{2015}) \bibinfo{pages}{171--193}.
%Type = Book
\bibitem[{Kalidindi(2015)}]{kalidindi2015hierarchical}
\bibinfo{author}{S.~R. Kalidindi}, \bibinfo{title}{Hierarchical materials
  informatics: novel analytics for materials data},
  \bibinfo{publisher}{Elsevier}, \bibinfo{year}{2015}.
%Type = Article
\bibitem[{Kim and Admal(2023)}]{JKIM:2023}
\bibinfo{author}{J.~Kim}, \bibinfo{author}{N.~C. Admal},
\newblock \bibinfo{title}{A stochastic framework for evolving grain statistics
  using a neural network model for grain topology transformations},
\newblock \bibinfo{journal}{Computational Materials Science}
  \bibinfo{volume}{199} (\bibinfo{year}{2023}) \bibinfo{pages}{111812}.
%Type = Article
\bibitem[{Abbruzzese and Lücke(1986)}]{ABBRUZZESE1986905}
\bibinfo{author}{G.~Abbruzzese}, \bibinfo{author}{K.~Lücke},
\newblock \bibinfo{title}{A theory of texture controlled grain growth—{I}.
  derivation and general discussion of the model},
\newblock \bibinfo{journal}{Acta Metallurgica} \bibinfo{volume}{34}
  (\bibinfo{year}{1986}) \bibinfo{pages}{905--914}.
%Type = Article
\bibitem[{Pande et~al.(2001)Pande, Masumura, and Marsh}]{Pande:2001}
\bibinfo{author}{C.~S. Pande}, \bibinfo{author}{R.~A. Masumura},
  \bibinfo{author}{S.~P. Marsh},
\newblock \bibinfo{title}{Stochastic analysis of two-dimensional grain growth},
\newblock \bibinfo{journal}{Philosophical Magazine A} \bibinfo{volume}{81}
  (\bibinfo{year}{2001}) \bibinfo{pages}{1229--1239}.
%Type = Article
\bibitem[{Esedo{\=g}lu and Otto(2015)}]{Esedoglu:2015}
\bibinfo{author}{S.~Esedo{\=g}lu}, \bibinfo{author}{F.~Otto},
\newblock \bibinfo{title}{Threshold dynamics for networks with arbitrary
  surface tensions},
\newblock \bibinfo{journal}{Communications on Pure and Applied Mathematics}
  \bibinfo{volume}{68} (\bibinfo{year}{2015}) \bibinfo{pages}{808--864}.
%Type = Article
\bibitem[{Eren et~al.(2022)Eren, Runnels, and Mason}]{Mason:2022}
\bibinfo{author}{E.~Eren}, \bibinfo{author}{B.~Runnels},
  \bibinfo{author}{J.~Mason},
\newblock \bibinfo{title}{Comparison of evolving interfaces, triple points, and
  quadruple points for discrete and diffuse interface methods},
\newblock \bibinfo{journal}{Computational Materials Science}
  \bibinfo{volume}{213} (\bibinfo{year}{2022}) \bibinfo{pages}{111632}.
%Type = Article
\bibitem[{Dziwnik et~al.(2017)Dziwnik, Münch, and Wagner}]{Dziwnik:2017}
\bibinfo{author}{M.~Dziwnik}, \bibinfo{author}{A.~Münch},
  \bibinfo{author}{B.~Wagner},
\newblock \bibinfo{title}{An anisotropic phase-field model for solid-state
  dewetting and its sharp-interface limit},
\newblock \bibinfo{journal}{Nonlinearity} \bibinfo{volume}{30}
  (\bibinfo{year}{2017}) \bibinfo{pages}{1465}.
%Type = Article
\bibitem[{Martine et~al.(2019)Martine, {Martine La Boissonière}, Choksi, and
  Esedo{\=g}lu}]{Esedoglu:2019}
\bibinfo{author}{G.~Martine}, \bibinfo{author}{{Martine La Boissonière}},
  \bibinfo{author}{R.~Choksi}, \bibinfo{author}{K.~B.~S. Esedo{\=g}lu},
\newblock \bibinfo{title}{Statistics of grain growth: Experiment versus the
  phase-field-crystal and {M}ullins models},
\newblock \bibinfo{journal}{Materialia} \bibinfo{volume}{6}
  (\bibinfo{year}{2019}) \bibinfo{pages}{100280}.
%Type = Article
\bibitem[{Elder et~al.(2007)Elder, Provatas, Berry, Stefanovic, and
  Grant}]{PFC}
\bibinfo{author}{K.~R. Elder}, \bibinfo{author}{N.~Provatas},
  \bibinfo{author}{J.~Berry}, \bibinfo{author}{P.~Stefanovic},
  \bibinfo{author}{M.~Grant},
\newblock \bibinfo{title}{Phase-field crystal modeling and classical density
  functional theory of freezing},
\newblock \bibinfo{journal}{Physical Review B} \bibinfo{volume}{75}
  (\bibinfo{year}{2007}) \bibinfo{pages}{064107}.
%Type = Article
\bibitem[{Peng et~al.(2022)Peng, Bhattacharya, Naghibzadeh, Kinderlehrer,
  Suter, Dayal, and Rohrer}]{Rohrer:2022}
\bibinfo{author}{X.~Peng}, \bibinfo{author}{A.~Bhattacharya},
  \bibinfo{author}{S.~K. Naghibzadeh}, \bibinfo{author}{D.~Kinderlehrer},
  \bibinfo{author}{R.~Suter}, \bibinfo{author}{K.~Dayal},
  \bibinfo{author}{G.~S. Rohrer},
\newblock \bibinfo{title}{Comparison of simulated and measured grain volume
  changes during grain growth},
\newblock \bibinfo{journal}{Physcial Review Materials} \bibinfo{volume}{6}
  (\bibinfo{year}{2022}) \bibinfo{pages}{033402}.
%Type = Article
\bibitem[{Niño and Johnson(2023)}]{NINO2023111879}
\bibinfo{author}{J.~D. Niño}, \bibinfo{author}{O.~K. Johnson},
\newblock \bibinfo{title}{Influence of grain boundary energy anisotropy on the
  evolution of grain boundary network structure during 3{D} anisotropic grain
  growth},
\newblock \bibinfo{journal}{Computational Materials Science}
  \bibinfo{volume}{217} (\bibinfo{year}{2023}) \bibinfo{pages}{111879}.
%Type = Article
\bibitem[{Bulatov et~al.(2014)Bulatov, Reed, and Kumar}]{BULATOV2014161}
\bibinfo{author}{V.~V. Bulatov}, \bibinfo{author}{B.~W. Reed},
  \bibinfo{author}{M.~Kumar},
\newblock \bibinfo{title}{Grain boundary energy function for fcc metals},
\newblock \bibinfo{journal}{Acta Materialia} \bibinfo{volume}{65}
  (\bibinfo{year}{2014}) \bibinfo{pages}{161--175}.
%Type = Article
\bibitem[{Salvador and Esedo{\=g}lu(2019{\natexlab{a}})}]{Salvador:2019}
\bibinfo{author}{T.~Salvador}, \bibinfo{author}{S.~Esedo{\=g}lu},
\newblock \bibinfo{title}{A simplified threshold dynamics algorithm for
  isotropic surface energies},
\newblock \bibinfo{journal}{Journal of Scientific Computing}
  \bibinfo{volume}{79} (\bibinfo{year}{2019}{\natexlab{a}})
  \bibinfo{pages}{648--669}.
%Type = Article
\bibitem[{Salvador and Esedo{\=g}lu(2019{\natexlab{b}})}]{Salvador:2019_2}
\bibinfo{author}{T.~Salvador}, \bibinfo{author}{S.~Esedo{\=g}lu},
\newblock \bibinfo{title}{The role of surface tension and mobility model in
  simulations of grain growth},
\newblock \bibinfo{journal}{arXiv:1907.11574}
  (\bibinfo{year}{2019}{\natexlab{b}}).
%Type = Article
\bibitem[{Valiev et~al.(2000)Valiev, Islamgaliev, and Alexandrov}]{ECAP:2000}
\bibinfo{author}{R.~Z. Valiev}, \bibinfo{author}{R.~K. Islamgaliev},
  \bibinfo{author}{I.~V. Alexandrov},
\newblock \bibinfo{title}{Bulk nanostructured materials from severe plastic
  deformation},
\newblock \bibinfo{journal}{Progress in Materials Science} \bibinfo{volume}{45}
  (\bibinfo{year}{2000}) \bibinfo{pages}{103--189}.
%Type = Article
\bibitem[{Semiatin et~al.(2004)Semiatin, Salem, and Saran}]{ECAP:2004}
\bibinfo{author}{S.~L. Semiatin}, \bibinfo{author}{A.~A. Salem},
  \bibinfo{author}{M.~J. Saran},
\newblock \bibinfo{title}{Models for severe plastic deformation by
  equal-channel angular extrusion},
\newblock \bibinfo{journal}{JOM} \bibinfo{volume}{56} (\bibinfo{year}{2004})
  \bibinfo{pages}{69--77}.
%Type = Article
\bibitem[{Zhu and Langdon(2004)}]{Langdon}
\bibinfo{author}{Y.~T. Zhu}, \bibinfo{author}{T.~G. Langdon},
\newblock \bibinfo{title}{The fundamentals of nanostructured materials
  processed by severe plastic deformation},
\newblock \bibinfo{journal}{JOM} \bibinfo{volume}{56} (\bibinfo{year}{2004})
  \bibinfo{pages}{58--63}.
%Type = Article
\bibitem[{Muñoz et~al.(2020)Muñoz, Bolmaro, Jr, Zhilyaev, and
  María}]{ECAP:2020}
\bibinfo{author}{J.~A. Muñoz}, \bibinfo{author}{R.~E. Bolmaro},
  \bibinfo{author}{A.~M.~J. Jr}, \bibinfo{author}{A.~Zhilyaev},
  \bibinfo{author}{J.~María},
\newblock \bibinfo{title}{Prediction of generation of high- and low-angle grain
  boundaries ({HAGB} and {LAGB}) during severe plastic deformation},
\newblock \bibinfo{journal}{Metallurgical and Materials Transactions A}
  \bibinfo{volume}{51} (\bibinfo{year}{2020}) \bibinfo{pages}{4674--4684}.
%Type = Article
\bibitem[{Merriman et~al.(1992)Merriman, Bence, and Osher}]{MBO}
\bibinfo{author}{B.~Merriman}, \bibinfo{author}{J.~K. Bence},
  \bibinfo{author}{S.~J. Osher},
\newblock \bibinfo{title}{Diffusion generated motion by mean curvature},
\newblock \bibinfo{journal}{Proceedings of the Computational Crystal Growers
  Workshop}  (\bibinfo{year}{1992}) \bibinfo{pages}{72--83}.
%Type = Article
\bibitem[{Zaitzeff et~al.(2020)Zaitzeff, Esedo{\=g}lu, and
  Garikipati}]{ZAITZEFF2020109404}
\bibinfo{author}{A.~Zaitzeff}, \bibinfo{author}{S.~Esedo{\=g}lu},
  \bibinfo{author}{K.~Garikipati},
\newblock \bibinfo{title}{Second order threshold dynamics schemes for two phase
  motion by mean curvature},
\newblock \bibinfo{journal}{Journal of Computational Physics}
  \bibinfo{volume}{410} (\bibinfo{year}{2020}) \bibinfo{pages}{109404}.
%Type = Article
\bibitem[{Elsey et~al.(2009)Elsey, Esedo{\=g}lu, and Smereka}]{Esedoglu:2009}
\bibinfo{author}{M.~Elsey}, \bibinfo{author}{S.~Esedo{\=g}lu},
  \bibinfo{author}{P.~Smereka},
\newblock \bibinfo{title}{Diffusion generated motion for grain growth in two
  and three dimensions},
\newblock \bibinfo{journal}{Journal of Computational Physics}
  \bibinfo{volume}{228} (\bibinfo{year}{2009}) \bibinfo{pages}{8015--8033}.
%Type = Article
\bibitem[{Elsey et~al.(2011)Elsey, Esedo{\=g}lu, and Smereka}]{Esedoglu:2011}
\bibinfo{author}{M.~Elsey}, \bibinfo{author}{S.~Esedo{\=g}lu},
  \bibinfo{author}{P.~Smereka},
\newblock \bibinfo{title}{Large scale simulations and parameter study for a
  simple recrystallization model},
\newblock \bibinfo{journal}{Philosophical Magazine} \bibinfo{volume}{91}
  (\bibinfo{year}{2011}) \bibinfo{pages}{1607--1642}.
%Type = Article
\bibitem[{Jacobs and Zhang(2017)}]{ESEDOLU201762}
\bibinfo{author}{S.~E.~M. Jacobs}, \bibinfo{author}{P.~Zhang},
\newblock \bibinfo{title}{Kernels with prescribed surface tension \& mobility
  for threshold dynamics schemes},
\newblock \bibinfo{journal}{Journal of Computational Physics}
  \bibinfo{volume}{337} (\bibinfo{year}{2017}) \bibinfo{pages}{62--83}.
%Type = Book
\bibitem[{Herring(1951)}]{Herring}
\bibinfo{author}{C.~Herring}, \bibinfo{title}{Surface Tension as a {M}otivation
  for {S}intering}, \bibinfo{publisher}{McGraw Hill}, \bibinfo{year}{1951}.
  \DOIprefix\doi{10.1007/978-3-642-59938-5_2}.
%Type = Article
\bibitem[{Ishii et~al.(1999)Ishii, Pires, and Souganidis}]{Threshold:viscosity}
\bibinfo{author}{H.~Ishii}, \bibinfo{author}{G.~E. Pires},
  \bibinfo{author}{P.~E. Souganidis},
\newblock \bibinfo{title}{Threshold dynamics type approximation schemes for
  propagating fronts},
\newblock \bibinfo{journal}{Journal of the MathematicalSociety of Japan}
  \bibinfo{volume}{51} (\bibinfo{year}{1999}) \bibinfo{pages}{267--308}.
%Type = Article
\bibitem[{Lawrence et~al.(2015)Lawrence, Rickman, Harmer, and
  Rollet}]{Rollet:2015}
\bibinfo{author}{A.~Lawrence}, \bibinfo{author}{J.~M. Rickman},
  \bibinfo{author}{M.~P. Harmer}, \bibinfo{author}{A.~D. Rollet},
\newblock \bibinfo{title}{Parsing abnormal grain growth},
\newblock \bibinfo{journal}{Acta Materialia} \bibinfo{volume}{103}
  (\bibinfo{year}{2015}) \bibinfo{pages}{681--687}.
%Type = Article
\bibitem[{Atkinson(1988)}]{Atkinson}
\bibinfo{author}{H.~V. Atkinson},
\newblock \bibinfo{title}{Overview no. 65: Theories of normal grain growth in
  pure single phase systems},
\newblock \bibinfo{journal}{Acta Metallurgica} \bibinfo{volume}{36}
  (\bibinfo{year}{1988}) \bibinfo{pages}{469--491}.
%Type = Article
\bibitem[{Mullins and Vi{\~n}als(2002)}]{Mullins:2002}
\bibinfo{author}{W.~Mullins}, \bibinfo{author}{J.~Vi{\~n}als},
\newblock \bibinfo{title}{Linear bubble model of abnormal grain growth},
\newblock \bibinfo{journal}{Acta Materialia} \bibinfo{volume}{50}
  (\bibinfo{year}{2002}) \bibinfo{pages}{2945--2954}.
%Type = Article
\bibitem[{Abbruzzese and Lücke(1986)}]{ABBRUZZESE}
\bibinfo{author}{G.~Abbruzzese}, \bibinfo{author}{K.~Lücke},
\newblock \bibinfo{title}{A theory of texture controlled grain growth—i.
  derivation and general discussion of the model},
\newblock \bibinfo{journal}{Acta Metallurgica} \bibinfo{volume}{34}
  (\bibinfo{year}{1986}) \bibinfo{pages}{905--914}.
%Type = Article
\bibitem[{Holm et~al.(2010)Holm, Olmsted, and Foiles}]{Holm:2010}
\bibinfo{author}{E.~A. Holm}, \bibinfo{author}{D.~L. Olmsted},
  \bibinfo{author}{S.~M. Foiles},
\newblock \bibinfo{title}{Comparing grain boundary energies in face-centered
  cubic metals: {A}l, {A}u, {C}u and {N}i},
\newblock \bibinfo{journal}{Scripta Materialia} \bibinfo{volume}{63}
  (\bibinfo{year}{2010}) \bibinfo{pages}{905--908}.
%Type = Article
\bibitem[{Bulatov et~al.(2014)Bulatov, Reed, and Kumar}]{Bulatov:2014}
\bibinfo{author}{V.~V. Bulatov}, \bibinfo{author}{B.~W. Reed},
  \bibinfo{author}{M.~Kumar},
\newblock \bibinfo{title}{Grain boundary energy function for fcc metals},
\newblock \bibinfo{journal}{Acta Materialla} \bibinfo{volume}{65}
  (\bibinfo{year}{2014}) \bibinfo{pages}{161--175}.
%Type = Article
\bibitem[{Wolf(1990)}]{Wolf:1990}
\bibinfo{author}{D.~Wolf},
\newblock \bibinfo{title}{Structure-energy correlation for grain boundaries in
  fcc metals---{III}. {S}ymmetrical tilt boundaries},
\newblock \bibinfo{journal}{Acta Metallugica et Materialia}
  \bibinfo{volume}{38} (\bibinfo{year}{1990}) \bibinfo{pages}{781--790}.
%Type = Article
\bibitem[{Lapovok et~al.(2015)Lapovok, Thomson, Cottam, and
  Estrin}]{Rupert:2014}
\bibinfo{author}{R.~Lapovok}, \bibinfo{author}{P.~F. Thomson},
  \bibinfo{author}{R.~Cottam}, \bibinfo{author}{Y.~Estrin},
\newblock \bibinfo{title}{Nanocrystalline grain boundary engineering:
  Increasing ${\Sigma}$3 boundary fraction in pure {Ni} with thermomechanical
  treatments},
\newblock \bibinfo{journal}{Acta Materialia} \bibinfo{volume}{86}
  (\bibinfo{year}{2015}) \bibinfo{pages}{43--54}.
%Type = Article
\bibitem[{Dao et~al.(2006)Dao, Lu, Shen, and Suresh}]{twinProperty}
\bibinfo{author}{M.~Dao}, \bibinfo{author}{L.~Lu}, \bibinfo{author}{Y.~F.
  Shen}, \bibinfo{author}{S.~Suresh},
\newblock \bibinfo{title}{Strength, strain-rate sensitivity and ductility of
  copper with nanoscale twins},
\newblock \bibinfo{journal}{Acta Materialia} \bibinfo{volume}{54}
  (\bibinfo{year}{2006}) \bibinfo{pages}{5421--5432}.
%Type = Article
\bibitem[{Bay et~al.(1992)Bay, Hansen, Hughes, and
  Kuhlmann-{W}ilsdorf}]{subgrain:1992}
\bibinfo{author}{B.~Bay}, \bibinfo{author}{N.~Hansen}, \bibinfo{author}{D.~A.
  Hughes}, \bibinfo{author}{D.~Kuhlmann-{W}ilsdorf},
\newblock \bibinfo{title}{Overview no. 96 evolution of f.c.c. deformation
  structures in polyslip},
\newblock \bibinfo{journal}{Acta Metallurgica et Materialia}
  \bibinfo{volume}{40} (\bibinfo{year}{1992}) \bibinfo{pages}{205--219}.
%Type = Article
\bibitem[{Lapovok et~al.(2005)Lapovok, Thomson, Cottam, and
  Estrin}]{subgrain:2005}
\bibinfo{author}{R.~Lapovok}, \bibinfo{author}{P.~F. Thomson},
  \bibinfo{author}{R.~Cottam}, \bibinfo{author}{Y.~Estrin},
\newblock \bibinfo{title}{The effect of grain refinement by warm equal channel
  angular extrusion on room temperature twinning in magnesium alloy {ZK}60},
\newblock \bibinfo{journal}{Journal of Materials Science} \bibinfo{volume}{40}
  (\bibinfo{year}{2005}) \bibinfo{pages}{1699--1708}.
%Type = Article
\bibitem[{Grand et~al.(2022)Grand, Flipon, Gaillac, and
  Bernacki}]{Bernacki:2022}
\bibinfo{author}{V.~Grand}, \bibinfo{author}{B.~Flipon},
  \bibinfo{author}{A.~Gaillac}, \bibinfo{author}{M.~Bernacki},
\newblock \bibinfo{title}{Simulation of continuous dynamic recrystallization
  using a level-set method},
\newblock \bibinfo{journal}{Materials} \bibinfo{volume}{15}
  (\bibinfo{year}{2022}) \bibinfo{pages}{8547}.
%Type = Article
\bibitem[{Julie et~al.(2021)Julie, Dash, Wasekar, David, and
  Kamruddin}]{David:2021}
\bibinfo{author}{S.~Julie}, \bibinfo{author}{M.~K. Dash},
  \bibinfo{author}{N.~P. Wasekar}, \bibinfo{author}{C.~David},
  \bibinfo{author}{M.~Kamruddin},
\newblock \bibinfo{title}{Effect of annealing and irradiation on the evolution
  of texture and grain boundary interface in electrodeposited nanocrystalline
  nickel of varying grain sizes},
\newblock \bibinfo{journal}{Surface \& Coatings Technology}
  \bibinfo{volume}{426} (\bibinfo{year}{2021}) \bibinfo{pages}{127770}.
%Type = Article
\bibitem[{Esedo{\=g}lu(2016)}]{Selim_Grotation}
\bibinfo{author}{S.~Esedo{\=g}lu},
\newblock \bibinfo{title}{Grain size distribution under simultaneous grain
  boundary migration and grain rotation in two dimensions},
\newblock \bibinfo{journal}{Computational Materials Science}
  \bibinfo{volume}{121} (\bibinfo{year}{2016}) \bibinfo{pages}{209--216}.
%Type = Article
\bibitem[{Admal et~al.(2018)Admal, Po, and Marian}]{Admal:2018}
\bibinfo{author}{N.~C. Admal}, \bibinfo{author}{G.~Po},
  \bibinfo{author}{J.~Marian},
\newblock \bibinfo{title}{A unified framework for polycrystal plasticity with
  grain boundary evolution},
\newblock \bibinfo{journal}{International Journal of Plasticity}
  \bibinfo{volume}{106} (\bibinfo{year}{2018}) \bibinfo{pages}{1--30}.
%Type = Article
\bibitem[{Wei et~al.(2020)Wei, Zhang, Han, Srolovitz, and
  Xiang}]{Srolovitz:2020}
\bibinfo{author}{C.~Wei}, \bibinfo{author}{L.~Zhang}, \bibinfo{author}{J.~Han},
  \bibinfo{author}{D.~J. Srolovitz}, \bibinfo{author}{Y.~Xiang},
\newblock \bibinfo{title}{Grain boundary triple junction dynamics: a continuum
  disconnection model},
\newblock \bibinfo{journal}{SIAM Journal on Applied Mathematics}
  \bibinfo{volume}{80} (\bibinfo{year}{2020}) \bibinfo{pages}{1101--1122}.
%Type = Article
\bibitem[{Runnels and Agrawal(2020)}]{Runnels:2020}
\bibinfo{author}{B.~Runnels}, \bibinfo{author}{V.~Agrawal},
\newblock \bibinfo{title}{Phase field disconnections: A continuum method for
  disconnection-mediated grain boundary motion},
\newblock \bibinfo{journal}{Scripta Materialia} \bibinfo{volume}{186}
  (\bibinfo{year}{2020}) \bibinfo{pages}{6--10}.

\end{thebibliography}

\end{document}